\documentclass[journal,engine=pdflatex]{IEEEtran}
\ifCLASSINFOpdf
\else
\fi
\usepackage{mcode}
\usepackage[linesnumbered,ruled]{algorithm2e}
\usepackage{algpseudocode}
\usepackage{setspace}
\usepackage{array}
\usepackage{url}
\usepackage{bm}
\usepackage{graphicx}
\usepackage{color}
\usepackage[caption=false]{subfig}
\usepackage{changes}
\usepackage{soul}

\usepackage{draftwatermark}          
\SetWatermarkText{Preprint}          
\SetWatermarkLightness{0.9}          
\SetWatermarkScale{1.1}              

\graphicspath{{figures/}}

\begin{document}

\title{Beyond Data-Physics Consistency: A Cross-Correlated Physics-Informed Neural Network for Robust Inverse Scattering}

\author{Shilong~Sun
\thanks{Preprint. Not peer-reviewed. This work has been submitted to the IEEE for possible publication. Copyright may be transferred without notice, after which this version may no longer be accessible.}
\thanks{S. Sun (emails: sunshilong@nudt.edu.cn) is with College of Electronic Science and Technology, National University of Defense Technology, Changsha, China}
\thanks{This work was supported by the National Natural Science Foundation of China under Grant 62471476 and 62231026. }
}

\maketitle

\begin{abstract}
    The electromagnetic inverse scattering problem (ISP), due to its inherent strong nonlinearity and severe ill-posedness, has long been a core challenge in microwave imaging. In recent years, physics-informed neural networks (PINNs) have provided a novel paradigm for solving ISPs by embedding Maxwell's equations into the deep learning optimization process. However, conventional PINN methods rely solely on independent data-equation and state-equation residuals to construct the consistency loss, which easily causes them to fall into local minima and suffer from low computational efficiency when facing high-contrast targets or multi-frequency observation data. To transcend the traditional data-physics consistency framework, this paper proposes a novel cross-correlated physics-informed neural network (CC-PINN). The core innovations of this work include: (1) constructing a Fourier feature MLP network architecture based on weight normalization, which possesses excellent capability for solving inverse scattering problems; (2) introducing a cross-correlated residual term that directly couples the reconstructed dielectric parameters and the predicted internal total field to the external observation field, breaking the decoupling between the contrast source and the permittivity optimization in traditional PINNs and significantly enhancing the robustness of PINNs for ISP; (3) introducing a zero-padding-based 2D-FFT acceleration algorithm, which reduces the computational complexity of the forward Green's function integration. Experimental results on synthetic and measured data demonstrate that CC-PINN can reconstruct high-contrast dielectric targets with high fidelity, and its convergence robustness far exceeds that of PINN algorithms using classical cost functions, regardless of whether simultaneous multi-frequency processing or frequency-hopping strategies are employed.
\end{abstract}


\section{Introduction}

    The core of the inverse scattering problem lies in retrieving physical parameters from wave equations, encompassing acoustic wave equations (widely used in geophysical acoustic exploration~\cite{Tarantola1984,Schuster2017}) and electromagnetic wave equations (widely used in medical imaging~\cite{Isaacson1986,Seo2003,Cheney2000}). The inversion theories and methods for both types of wave equations are universally applicable. Electromagnetic forward problems, due to their inherent complexity, generally require substantial computational resources, making the computational complexity and solution difficulty of electromagnetic inverse scattering problems significantly higher, which stems from the ill-posedness of the problem and its nonlinear dependence on the target parameters~\cite{colton2012inverse}.

    In the field of electromagnetic inverse scattering, quantitative inversion aims to reconstruct the target's dielectric parameters (equivalent to the contrast function described later) from limited electromagnetic field measurement data. The core challenge lies in effectively addressing the inherent ill-posedness and nonlinearity. Electromagnetic quantitative inversion methods can be classified into three categories based on their principles: linear quantitative inversion methods, iterative optimization-based methods, and neural-network-driven methods. A representative of linear quantitative inversion is diffraction tomography~\cite{Malcolm_Slaney,851784,10144396}, which utilizes the Born or Rytov approximation to establish a linear mapping between the target dielectric parameters and the spatial spectrum and then achieves quantitative reconstruction via Fourier transform. However, this class of methods is valid only under the weak scattering assumption and is applicable only to low-contrast targets. Iterative optimization-based methods handle the nonlinearity and ill-posedness of inverse scattering through an iterative optimization framework, achieving high reconstruction accuracy in general scenarios, albeit with higher computational complexity. The contrast source inversion (CSI) algorithm~\cite{van1997contrast} is a pioneering work of this class, which has been followed by the development of the subspace optimization method (SOM)~\cite{Li2010,5210141,10663369} and cross-correlated CSI (CC-CSI)~\cite{sun2017cross,sun2018inversion,11106328}. Another representative method is the distorted Born iterative method~\cite{chew1990reconstruction}. Unlike local optimization methods, stochastic-optimization-based inverse scattering techniques avoid local optima through random exploration, are suitable for high-contrast inversion, and offer high accuracy~\cite{HajebiTGRS2022,hajebi2024subsurface,PastorinoTAP2007}, but they incur high computational cost and their convergence is sensitive to hyperparameters. To the best of our knowledge, S.~Caorsi and P.~Gamba were the first to apply neural networks to electromagnetic inverse scattering problems~\cite{752198}. In recent years, the explosive development of artificial intelligence has propelled the application of convolutional neural networks (CNNs)~\cite{YuSun18,8565987,10028725,DastfalTGRSL2025,10666745}, generative neural networks~\cite{9852109,9897092}, and attention mechanisms~\cite{10572305} in electromagnetic inverse scattering. These methods naturally possess the real-time inference capability of neural networks, making them highly attractive in practical applications compared to traditional time-consuming inversion algorithms. However, establishing a direct mapping from scattered field data to target contrast is extremely difficult, forcing these methods to heavily rely on massive and expensive paired training datasets, resulting in concerning generalization ability in practical applications.

    Against this background, the introduction of physics-informed neural networks (PINNs)~\cite{RAISSI2019686} has opened a new avenue for unsupervised ISP inversion. Early ISP-PINN works (e.g., Chen et al.~\cite{Chen2019PhysicsinformedNN}) utilized neural networks to implicitly represent the spatial medium distribution and employed the data equation and state equation, both based on integral equations, as unsupervised loss functions. Although preliminary results were achieved, existing PINN architectures suffer from two fatal flaws: first, conventional loss functions pursue only independent data-physics consistency and lack a cross-validation mechanism, making them extremely prone to falling into local optima for strong scatterers; second, each forward propagation of PINNs requires computing dense Green's function integrals, causing the training time to increase exponentially. To address these challenges, this paper proposes a cross-correlated physics-informed neural network (CC-PINN) inversion algorithm that transcends the traditional consistency framework. First, this paper designs and constructs a Fourier feature network architecture based on weight normalization: weight normalization decouples the norm and direction of the weights, allowing the network training process to automatically balance multiple tasks and preventing a specific loss from dominating the training~\cite{10.5555/3157096.3157197}. The introduction of multi-scale random Fourier features endows the network with strong expressive power in each frequency band~\cite{DBLP:conf/nips/RahimiR07,NEURIPS2020_55053683}. Second, inspired by the traditional CC-CSI algorithm~\cite{sun2017cross}, this paper introduces a cross-correlation term for the first time into neural network optimization and constructs a cross-correlated cost function suitable for the PINN inversion architecture, enabling the network to achieve both good robustness and high inversion accuracy. Finally, by constructing a 2D-FFT operation that is strictly equivalent to spatial linear convolution, the last barrier to PINN computational efficiency is removed. Validation with synthetic and measured data for transverse magnetic (TM) waves demonstrates that the CC-PINN framework significantly enhances the convergence robustness of the network in solving electromagnetic inverse scattering problems. This framework is applicable to various network optimization architectures based on the data-physics consistency principle and provides universal guidance for more effectively utilizing data-physics consistency in quantitative inversion.

    The remainder of this paper is organized as follows: Section~\ref{sec.ProSta} establishes the mathematical model of the two-dimensional inverse scattering problem and separately describes the weight-normalization-based Fourier feature network architecture, the cross-correlated cost function construction method, and the zero-padding 2D-FFT-based forward acceleration technique. In Section~\ref{sec.Validation}, a comprehensive performance analysis of the proposed algorithm is carried out using numerical simulations and measured data, verifying the superior convergence robustness of the proposed CC-PINN inversion framework. Finally, Section~\ref{sec.Conclusion} summarizes the main contributions, the limitations and future research directions.

\section{Mathematical Modeling and Algorithm Description}\label{sec.ProSta}

\subsection{Modeling of the Electromagnetic Inverse Scattering Problem}

    Consider a typical two-dimensional TM polarized inverse scattering measurement scenario. Assuming a time-harmonic factor $e^{j\omega t}$, a bounded non-magnetic unknown target is located within the domain of interest (DOI) $D$. For the multi-frequency broadband sensing configuration, let the set of discrete incident wave frequencies be $\{f_1, f_2, \dots, f_I\}$, with corresponding angular frequencies $\omega_i = 2\pi f_i$. Under single-frequency $\omega_i$ illumination, the complex contrast function of the target at spatial position $\mathbf{r} \in D$ is defined in a frequency-dependent form:
    \begin{equation}
    \chi_i(\mathbf{r}) = \epsilon_r(\mathbf{r}) - 1 - j\frac{\sigma(\mathbf{r})}{\omega_i \epsilon_0}
    \end{equation}
    where $\epsilon_r(\mathbf{r})$ and $\sigma(\mathbf{r})$ are the unknown spatial distributions of relative permittivity and conductivity, respectively; $\epsilon_0$ is the permittivity of background free space.

    \begin{figure}[!t]
        \centering
        \includegraphics[width=0.5\linewidth]{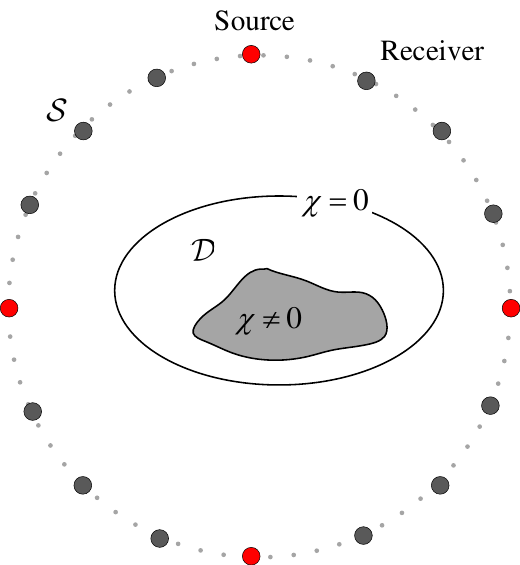}
        \caption{Geometric schematic diagram of the detection scenario for the two-dimensional electromagnetic inverse scattering problem.}
        \label{fig:configuration}
    \end{figure}

    This paper adopts a full-angle circular array observation configuration. $P$ transmitting antennas and $Q$ receiving antennas are uniformly deployed on the observation circle $S_{obs}$ outside the DOI $D$. Let the $p$-th transmitting antenna be located at $\mathbf{r}_p^t$ ($p=1, 2, \dots, P$), which generates a background incident electric field $E_{\text{inc}, i}^p(\mathbf{r})$ inside the region $D$ at frequency $f_i$. According to the volume equivalence principle, the inhomogeneity inside the target is equivalent to a contrast source induced by the incident wave: $J_i^p(\mathbf{r}) = \chi_i(\mathbf{r}) E_{\text{tot}, i}^p(\mathbf{r})$, where $E_{\text{tot}, i}^p(\mathbf{r})$ is the total internal electric field. The forward physical process of inverse scattering is governed by two sets of Lippmann-Schwinger integral equations with dual subscripts of frequency and transmitting antenna:
    \begin{itemize}
        \item State Equation: describes the self-consistent physical conservation relation of the internal total field
            \begin{equation}
            E_{\text{tot}, i}^p(\mathbf{r}) = E_{\text{inc}, i}^p(\mathbf{r}) + k_{0,f}^2 \int_D G_i(\mathbf{r}, \mathbf{r}') J_i^p(\mathbf{r}') d\mathbf{r}', \quad \mathbf{r}, \mathbf{r}' \in D
            \end{equation}
            where $k_{0,f} = \omega_i / c$ is the free-space wavenumber and $c$ is the speed of light. Based on the time-harmonic convention, the two-dimensional free-space Green's function is expressed as $G_i(\mathbf{r}, \mathbf{r}') = -\frac{j}{4}H_0^{(2)}(k_{0,f}|\mathbf{r}-\mathbf{r}'|)$. To facilitate expression in subsequent deep network architectures, for the system covering all $P$ excitations, we abbreviate it in operator-matrix form as: $\mathbf{E}_{\text{tot}, i} = \mathbf{E}_{\text{inc}, i} + \mathcal{G}_{D, i} \mathbf{J}_i$.

        \item Data Equation: describes the external scattered field $E_{\text{meas}, i}^{p, q}$ generated at the $q$-th receiving antenna position $\mathbf{r}_q^r \in S_{obs}$ under the $p$-th excitation
            \begin{equation}
            E_{\text{meas}, i}^{p, q} = k_{0,f}^2 \int_D G_i(\mathbf{r}_q^r, \mathbf{r}') J_i^p(\mathbf{r}') d\mathbf{r}', \quad \mathbf{r}_q^r \in S_{obs}, \mathbf{r}' \in D
            \end{equation}
            Similarly, the entire set of excitation-receiver scattered data can be expressed in operator-matrix form: $\mathbf{E}_{\text{meas},i} = \mathcal{G}_{S, i} \mathbf{J}_i$.
    \end{itemize}
    The inverse scattering problem is to retrieve the frequency-independent intrinsic medium distributions $\epsilon_r(\mathbf{r})$ and $\sigma(\mathbf{r})$ given the multi-frequency observed dataset $\mathbf{E}_{\text{meas},i}$.

\subsection{Weight-Normalization-Based Fourier Feature Network Architecture}

    To map the two-dimensional spatial coordinates $\mathbf{r} = (x, y)$ into a continuous medium distribution, we implicitly parameterize the contrast function using a multilayer perceptron (MLP) network. However, standard MLPs generally exhibit the ``spectral bias'' phenomenon in deep learning theory, i.e., the network tends to preferentially fit low-frequency, smooth function mappings, while showing extremely poor learning ability for high-frequency features (such as sharp physical boundaries of targets and local high-contrast jumps), which is unacceptable in microwave inverse scattering that requires fine imaging.

\subsubsection{Fourier Feature Mapping}

    To overcome the spectral bias of standard MLPs in learning high-frequency spatial features, we introduce Fourier feature mapping at the input, forcibly projecting the low-dimensional coordinate $\mathbf{r}$ into a high-frequency periodic feature space:
    \begin{equation}
    \mathbf{\gamma}(\mathbf{r}) = \left[\,\sin(2\pi \mathbf{B}\mathbf{r}),\; \cos(2\pi \mathbf{B}\mathbf{r})\,\right]
    \label{eq:fourier_feature}
    \end{equation}
    The meaning and implementation details of each symbol in Equation~\eqref{eq:fourier_feature} are as follows:
    \begin{itemize}
        \item $\mathbf{r} = (x, y)\in\mathcal{D}$: two-dimensional spatial coordinate in meters. Before being fed into the network, the coordinates are normalized to the $[-1,1]$ interval, i.e., assuming the inversion region is $\text{ROI} = [-R, R]\times[-R, R]$, then $x_{\text{norm}} = x / R$, $y_{\text{norm}} = y / R$.
        \item $\mathbf{B}$: Gaussian random projection matrix of shape $2 \times m$, where $m$ is the mapping dimension. The elements of the matrix are independently and identically sampled from a Gaussian distribution $\mathcal{N}(0, \sigma^2)$, where the standard deviation $\sigma$ is a crucial hyperparameter in Fourier feature networks. A smaller $\sigma$ leads to smoother inversion results, causing loss of detail and blurred edges; a larger $\sigma$ means higher spatial frequencies are contained in the Fourier features, potentially resulting in granular artifacts in the inversion image. $\mathbf{B}$ remains fixed (non-trainable) during training, serving to provide the network with a fixed set of high-frequency orthogonal bases.
        \item $\mathbf{B}\mathbf{r}$: linearly transforms the $2$-dimensional coordinate into an $m$-dimensional space, where each dimension acquires an independent frequency component.
        \item $2\pi \mathbf{B}\mathbf{r}$: multiplies the projected result by $2\pi$, so that the phase of the sine/cosine functions covers multiple oscillation cycles, thereby producing rich high-frequency components.
        \item $\mathbf{\gamma}(\mathbf{r})$: the final output Fourier feature vector of length $2m$ (concatenation of sine and cosine). It maps the smooth low-dimensional coordinates into high-frequency, periodic features, forcing the network to receive high-frequency information at the input layer and effectively overcoming spectral bias.
    \end{itemize}

    The neural tangent kernel (NTK) of a standard MLP is essentially a low-pass filter, making it difficult for the network to learn high-frequency variations in the data. Fourier features explicitly inject a set of basis functions with random frequencies, which is equivalent to embedding the input space into a high-dimensional oscillatory feature space, after which the subsequent MLP only needs to learn linear combinations of these bases. The isotropy of the random Gaussian matrix $\mathbf{B}$ ensures uniform frequency coverage in all spatial directions, and keeping it fixed and non-trainable preserves the convexity of the problem and avoids overfitting.

    \begin{figure*}[!t]
        \centering
        \includegraphics[width=0.975\linewidth]{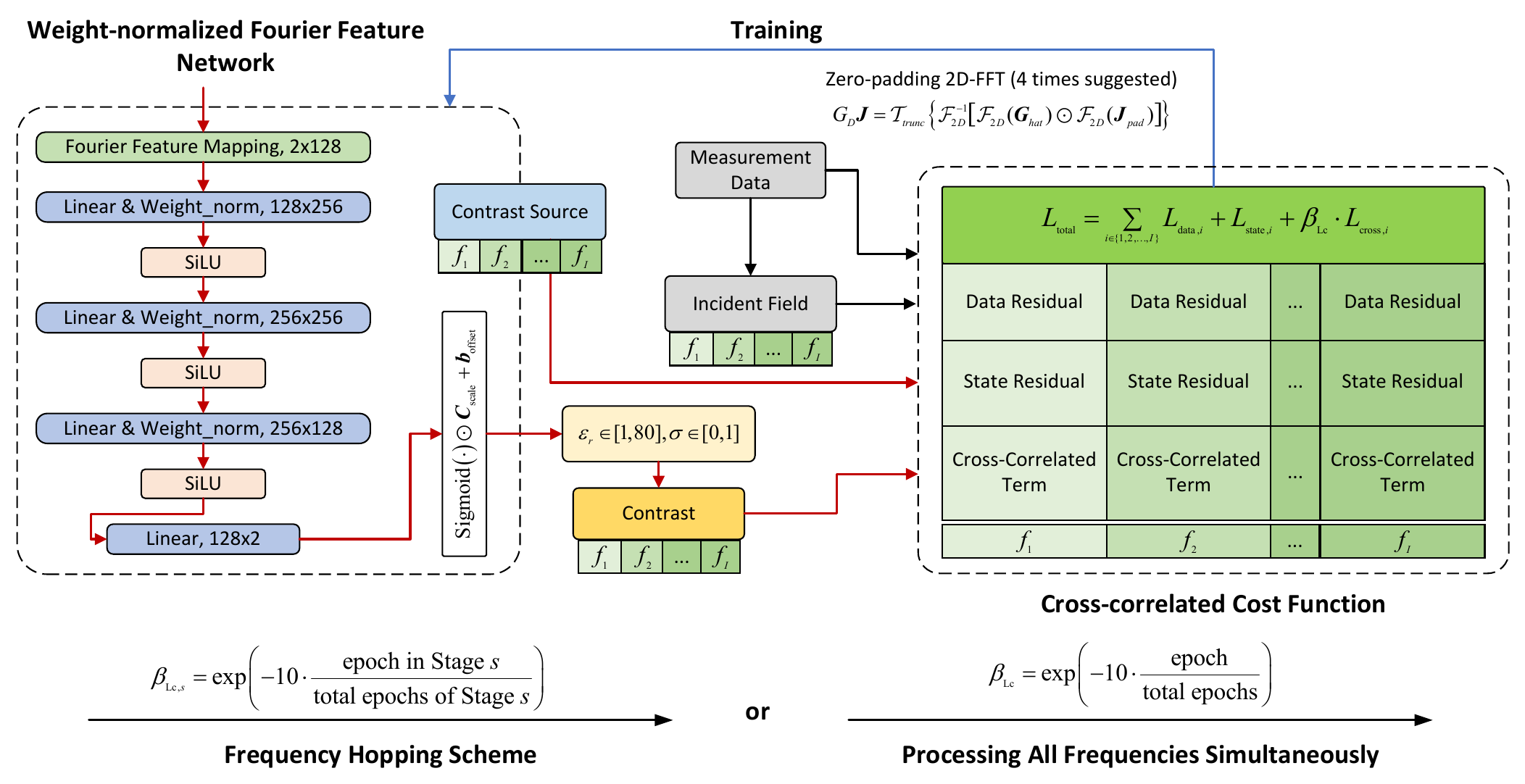}
        \caption{Architecture diagram of CC-PINN with weight-normalization-based Fourier feature network and zero-padding 2D-FFT acceleration.}
        \label{fig:Algorithm_Block_Diagram}
    \end{figure*}

\subsubsection{Weight-Normalized MLP Mapping}

    On top of the Fourier features, we construct a multilayer perceptron using weight normalization (proposed by Salimans \& Kingma (2016)~\cite{10.5555/3157096.3157197}), with the activation function $\text{SiLU}(\cdot)$,
    \begin{equation}
        \text{SiLU}(x) = x \cdot \sigma(x) = \frac{x}{1 + e^{-x}}
    \end{equation}
    where $\sigma(x)$ is the standard sigmoid function. The smoothness of SiLU avoids the gradient discontinuities introduced by the non-smooth turning point of ReLU at zero, helping to preserve the smoothness of the reconstructed image. The network ultimately outputs the relative permittivity and conductivity:
    \begin{equation}
    [\epsilon_r(\mathbf{r}),\; \sigma(\mathbf{r})] = \text{Sigmoid}\left( \text{MLP}_{\text{WNorm}}(\gamma(\mathbf{r}); \theta) \right) \;\odot\; \mathbf{C}_{\text{scale}} + \mathbf{b}_{\text{offset}}
    \label{eq:wnorm_output}
    \end{equation}
    The symbol definitions and implementation correspondences are as follows:
    \begin{itemize}
        \item $\text{MLP}_{\text{WNorm}}(\cdot;\theta)$: an MLP employing weight normalization. Weight normalization re-parameterizes each weight matrix $\mathbf{W}$ as $\mathbf{W} = g \cdot \frac{\mathbf{v}}{\|\mathbf{v}\|_2}$, where $g$ is a learnable scalar magnitude and $\mathbf{v}$ is a learnable direction vector. This operation decouples the direction and magnitude of the weights, making the optimization process more stable.
        \item $\theta$: all trainable parameters in the network, including $g$, $\mathbf{v}$ for each weight-normalized layer, and the weight and bias of the final layer (without weight normalization).
        \item $\text{Sigmoid}(\cdot)$: element-wise logistic function that compresses the network's raw output to the $(0,1)$ interval, ensuring non-negativity and boundedness of the physical quantities.
        \item $\mathbf{C}_{\text{scale}} = [C_\epsilon, C_\sigma]$: vector of scaling coefficients. In this paper, $C_\epsilon = 79$, $C_\sigma = 1$.
        \item $\mathbf{b}_{\text{offset}} = [b_\epsilon, b_\sigma]$: vector of offsets. In this paper, $b_\epsilon = 1$, $b_\sigma = 0$, so that the background medium ($\epsilon_r=1,\sigma=0$) corresponds to a sigmoid output close to $0$.
        \item $\epsilon_r(\mathbf{r})$: reconstructed relative permittivity distribution, dimensionless, ranging $1\sim80$.
        \item $\sigma(\mathbf{r})$: reconstructed conductivity distribution, in S/m, ranging $0\sim1$.
    \end{itemize}

    Fig.~\ref{fig:Algorithm_Block_Diagram} shows the overall structure of the network. The network structure is: input dimension $2m$ $\to$ linear layer (weight normalization) + SiLU $\to$ linear layer (weight normalization) + SiLU $\to$ linear layer (weight normalization) + SiLU $\to$ output linear layer (no weight normalization), with an output dimension of $2$. To facilitate the academic community in reproducing the algorithm, the core hyperparameters used in the simulation experiments are listed as follows:
    \begin{itemize}
        \item Grid division: In simulations, the domain range is $\text{ROI} = [-R, R]\times[-R, R]$, with $R = 0.5\,\text{m}$, divided into an $N = 64 \times 64$ grid.
        \item Network dimensions: The Fourier feature mapping dimension is set to $2m = 128$, with the Gaussian random projection matrix sampled from a Gaussian distribution with standard deviation $\sigma=1.0$; the number of neurons in the hidden layers is $[256, 256, 128, 2]$ in order.
        \item Initialization: The background is set as constant ($\epsilon_r=1.0, \sigma=0.0$), and the bias of the final network layer is initialized to $-3.0$ to ensure physical validity. The parameter $\mathbf{J}$ is initially estimated using back-propagation.
        \item Optimizer and learning rate: The Adam optimizer is used jointly, with a learning rate of $10^{-3}$ for the permittivity network and $2 \times 10^{-3}$ for the contrast source variable, accompanied by a CosineAnnealingLR decay schedule.
    \end{itemize}

    In the subsequent code implementation, the bias of the final linear layer is initialized to $-3.0$ (in fact, this initialization has little effect on the experimental results), and the weights adopt small random numbers with a mean of $0$ and a standard deviation of $10^{-3}$.

\subsection{Construction of the Cross-Correlated Cost Function}

    In the CC-PINN architecture, the contrast distribution $\boldsymbol{\chi}_i$ is the direct result of the forward inference of the network weights $\theta$, while the internal contrast source matrix $\mathbf{J}_i$ is decoupled within the PyTorch computational graph and defined as explicit complex-valued learnable leaf node tensors. These two types of variables are jointly optimized within the same back-propagation computational graph.

    For a single frequency $f_i$, we construct a physics-informed loss function consisting of three core terms:
    \begin{equation}
    \begin{aligned}
    L_{\text{data}, i} &= \frac{||\mathcal{G}_{S, i} \mathbf{J}_i - \mathbf{E}_{\text{meas},i}||_F^2}{||\mathbf{E}_{\text{meas},i}||_F^2} \\
    L_{\text{state}, i} &= \frac{||\boldsymbol{\chi}_i(\theta) \odot (\mathbf{E}_{\text{inc}, i} + \mathcal{G}_{D, i} \mathbf{J}_i) - \mathbf{J}_i||_F^2}{||\mathbf{E}_{\text{inc}, i}||_F^2} \\
    L_{\text{cross}, i} &= \frac{||\mathcal{G}_{S, i} \big(\boldsymbol{\chi}_i(\theta) \odot (\mathbf{E}_{\text{inc}, i} + \mathcal{G}_{D, i} \mathbf{J}_i)\big) - \mathbf{E}_{\text{meas},i}||_F^2}{||\mathbf{E}_{\text{meas},i}||_F^2}
    \end{aligned}
    \end{equation}
    where $\odot$ denotes the element-wise Hadamard product of matrices, and $||\cdot||_F$ represents the Frobenius norm. The cross-correlation term $L_{\text{cross}, i}$ directly maps the network-inferred $\boldsymbol{\chi}_i$ to the far-field observation end, forming a closed-loop physical verification that significantly constrains the solution space of the nonlinear search.

    The reader may notice that in the denominators of the state term $L_{\text{state}, i}$ and the cross term $L_{\text{cross}, i}$, we consistently use constant invariant physical quantities (the background incident field norm $||\mathbf{E}_{\text{inc}, i}||_F^2$ and the observed scattered field norm $||\mathbf{E}_{\text{meas},i}||_F^2$), rather than following certain traditional numerical algorithms that use the estimated norm at the current iteration (e.g., $||\boldsymbol{\chi}_i \mathbf{E}_{\text{tot}, i}||_F^2$). This design is crucial for the optimization stability of PINNs. In neural network training based on automatic differentiation, if the denominator contains variables involving network weights $\theta$ or $\mathbf{J}_i$, the network can easily discover a ``trivial solution'' loophole: the network does not drive the numerator toward zero by fitting the physical equations, but instead abnormally amplifies the values of $\boldsymbol{\chi}_i$ or $\mathbf{J}_i$, infinitely increasing the denominator, thereby mathematically ``cheating'' to reduce the overall loss. By locking the normalization denominators as definite, gradient-free physical constant matrix norms, CC-PINN provides an absolute, non-drifting physical anchor for global gradient descent, forcing the network to complete optimization solely by reducing the physical residual numerators.

    On one hand, the Cross Term forces the contrast currently output by the network to consistently conform to the physical scattering process from end to end and match the external observations, effectively enhancing algorithmic robustness. On the other hand, introducing the Cross Term essentially substitutes the state equation into the data equation. Because the contrast and contrast source are updated simultaneously, this increases the ``ruggedness'' of the loss landscape (introducing more local minima) to some extent, meaning that the improvement in robustness often comes at the cost of sacrificing part of the inversion accuracy. To address this issue, this paper proposes constructing a cross-correlated cost function with an exponential decay factor, i.e., defining the multi-frequency total loss function as
    \begin{equation}
        L_\text{total} = \sum_{i \in \{1, 2, \dots, I\}}  L_{\text{data}, i} + L_{\text{state}, i} + \beta_\text{Lc}\cdot L_{\text{cross}, i} 
    \end{equation}
    where the exponential decay factor $\beta_\text{Lc}$ is defined as
    \begin{equation}
        \beta_\text{Lc} = \exp\left(-10\cdot \frac{\text{epoch}}{\text{epochs}}\right)
    \end{equation}
    The introduction of the exponential decay factor $\beta_\text{Lc}$ balances the algorithmic robustness brought by the cross-correlation residual with the inversion accuracy of the traditional cost function.

    There are usually two strategies for processing multi-frequency data: processing all frequency points simultaneously, and the frequency-hopping strategy. Under the frequency-hopping strategy, we define the exponential decay factor $\beta_{\text{cross},s}$ in Stage $s$ as
    \begin{equation}
        \beta_{\text{cross},s} = \exp\left(-10\cdot \frac{\text{epoch in Stage } s}{\text{epochs of Stage } s}\right)
    \end{equation}
    Training neural networks is mostly based on parameter optimization algorithms such as Adam, and the transition when switching frequencies is often not as smooth as in traditional gradient-based optimization inversion methods. When processing different types of data, one often faces the dilemma of choosing between simultaneous multi-frequency processing and the frequency-hopping strategy. Based on our experience with synthetic and measured data in this paper, which strategy to use should be decided according to the data type, the characteristics of the inversion algorithm, the inversion difficulty of the target, and the specific situation, following the principle of optimizing inversion performance.

\subsection{Forward Acceleration Based on Zero-Padding 2D-FFT}

    In each iteration, the computation of $L_{\text{state}}$ and $L_{\text{cross}}$ requires the intensive calculation of the integral term $G_D \mathbf{J}$. Assume the ROI is discretized into an $N \times N$ grid (total number of grid cells $N_g = N^2$). In traditional computation, $G_D$ manifests as an enormous $N_g \times N_g$ dense complex matrix, and the computational cost of a single direct matrix-vector multiplication is $\mathcal{O}(N_g^2) = \mathcal{O}(N^4)$. When the grid is refined (for example, when $N=64$, $N^4 \approx 1.67 \times 10^7$ complex multiply-add operations), this computational cost becomes a fatal bottleneck for the PINN algorithm as the number of transmitting antennas $N_{tx}$ and frequency points increases.

    Note that the Green's function $G(\mathbf{r}, \mathbf{r}')$ exhibits translational invariance on a uniform grid, meaning that this integral is essentially a two-dimensional linear convolution. Drawing on the fast computation method in~\cite{10044704}, this paper introduces zero-padding during the network training process, expanding the original $N \times N$ domain $\mathbf{J}$ into a tensor $\mathbf{J}_{pad}$ of size $M_\text{pad}N \times M_\text{pad}N$, and constructing the Green's function kernel as a periodic kernel $\mathbf{G}_{hat}$ of the same dimension. Based on the circular convolution theorem, the linear convolution is losslessly equivalently transformed into:
    \begin{equation}
    G_D \mathbf{J} = \mathcal{T}_{trunc} \left\{ \mathcal{F}_{2D}^{-1} \Big[ \mathcal{F}_{2D}(\mathbf{G}_{hat}) \odot \mathcal{F}_{2D}(\mathbf{J}_{pad}) \Big] \right\}
    \end{equation}
    where $\mathcal{F}_{2D}$ denotes the two-dimensional FFT, $\odot$ is element-wise multiplication, and $\mathcal{T}_{trunc}$ is the operator that truncates the effective $N \times N$ region.

    The theoretical complexity of this method is reduced from $\mathcal{O}(N^4)$ to $\mathcal{O}(N^2\log N)$. Combined with the GPU's cuFFT library, the practical gains are even higher. In implementation, it should be noted that: the wavenumber $k_0$ differs for each frequency point, so the constant tensor $\mathcal{F}_{2D}(g_{\text{pad}})$ should be pre-computed and stored; zero-padding must be performed in both dimensions by adding $\left(M_\text{pad}-1\right)N$ zeros, ensuring that the padded region is located at the bottom-right to make it equivalent to linear convolution; all operations use single-precision complex numbers (complex64) to balance accuracy and memory. Considering the impact of spectral truncation on computational accuracy, 4$\times$ zero-padding is adopted in the simulations. Empirically, $M_\text{pad}=4$ provides sufficient accuracy with minimal computational overhead. FFT acceleration completely eliminates the computational bottleneck of internal integration, making real-time reconstruction of fine-grid, multi-frequency, multi-static data possible.

\section{Experimental Validation and Performance Analysis}\label{sec.Validation}

    In this section, using synthetic and measured data respectively, and based on the exact same weight-normalized Fourier feature network architecture, network parameters, and initialization strategy, we test and analyze the performance of the proposed CC-PINN inversion framework against the traditional cost function construction method (without the cross-correlation term). Both frequency-hopping and simultaneous multi-frequency processing strategies are considered in this paper. Regarding the impact of PINN initialization randomness on inversion results, all experiments are repeated over 11 independent runs, and the mean PSNR curves and boxplots commonly used in statistics are employed to comprehensively test the robustness and inversion accuracy of the proposed CC-PINN inversion framework, avoiding misleading conclusions caused by specific experimental cases.

\subsection{Synthetic Experiment Analysis}

\subsubsection{Simulation Scenario and Parameter Configuration}

\begin{figure}[!t]
        \centering
        \includegraphics[width=0.65\linewidth]{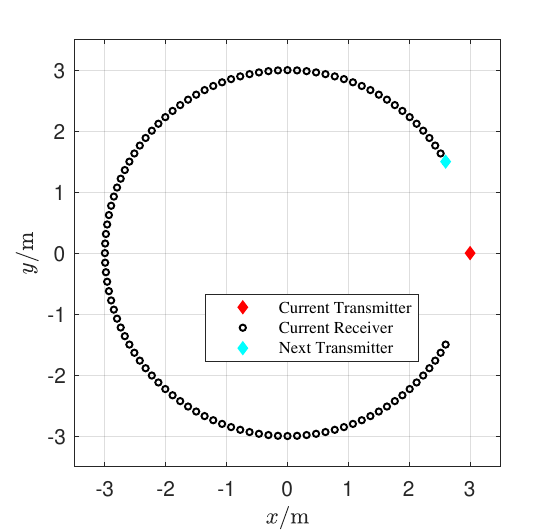}
        \caption{Probing configuration of the simulated data.}
        \label{fig:config}   
    \end{figure}

    The experimental configuration simulates a Fresnel data acquisition setup (as shown in Fig.~\ref{fig:config}), where transmitting and receiving elements are uniformly distributed on a concentric circle with a radius of 3~m. For each transmitting element, receiver data within a $30^\circ$ angular sector centered on it are excluded, retaining only measurement data within the remaining $300^\circ$ angular range with an angular step of $3^\circ$. Therefore, the measurement data dimension for TM-polarized inversion is $12 \times 101 \times N_\text{f}$. The simulation environment applies perfectly matched layers (PML) along the $x$ and $y$ boundaries to emulate an anechoic chamber, while periodic boundary conditions (PBC) are applied along the $z$ direction to maintain the two-dimensional configuration. The TM-polarized incident wave is generated by line sources aligned with the $z$-axis. To generate the scattered data, a uniform grid of $5~\text{mm} \times 5~\text{mm}$ is used, satisfying the criterion $\Delta \leq \lambda_0/15/\sqrt{\varepsilon_\text{r}}$~\cite{W.Shin2013}, where $\lambda_0$ denotes the free-space wavelength. The scattered field is obtained by subtracting the incident field from the total field. This paper employs the widely adopted ``Austria'' benchmark problem~\cite{belkebir1996using,litman1998reconstruction,van2001contrast,van2003multiplicative}. Specifically, the target to be reconstructed consists of two disks and one ring. In the defined coordinate system, the $z$-axis aligns with the axial direction of the objects: two disks, both of radius 0.1~m, with centers located at (0.3, $-0.15$)~m and (0.3, $0.15$)~m respectively; a ring centered at ($-0.1$, 0)~m, with outer radius 0.3~m and inner radius 0.15~m. It should be emphasized that the inversion of the ``Austria'' profile is widely considered a highly challenging inverse scattering problem in the existing literature.

    To analyze the inversion performance of the algorithm under different noise levels, complex white Gaussian noise is added to the total field of the observed data. The mathematical formula for noise addition is as follows:
    \begin{equation}
        \bm{y}'_{p,i} = \bm{y}_{p,i} + \sqrt{\frac{\|\bm{y}_{p,i}\|_2^2}{N \cdot 10^{\text{snr}/10}}} \cdot \frac{n_{\text{real}} + \text{j}\cdot n_{\text{imag}}}{\sqrt{2}}
    \end{equation}
    where $N$ denotes the length of $\bm{y}_{p,i}$; $n_{\text{real}}$ and $n_{\text{imag}}$ follow the standard normal distribution, i.e., $n_{\text{real}}, n_{\text{imag}} \sim \mathcal{N}(0, 1)$. The symbol ``j'' denotes the imaginary unit in complex numbers. Using noise-free data, three noisy data files at different signal-to-noise ratios (SNR $=20$ dB, 10 dB, and 0 dB) are generated and fixed, to ensure that the randomness of data noise can be excluded in algorithm comparison and performance analysis.

    The synthetic data includes three frequency points: $0.3$ GHz, $0.4$ GHz, and $0.5$ GHz. This paper separately tests and demonstrates two multi-frequency processing strategies: frequency-hopping and simultaneous multi-frequency processing. In the frequency-hopping strategy, the total number of epochs is set to 15000, Stage 1 ($0.3$ GHz) and Stage 2 ($0.3$ GHz $+$ $0.4$ GHz) equally share $40\%$ of the total epochs, and Stage 3 ($0.3$ GHz $+$ $0.4$ GHz $+$ $0.5$ GHz) accounts for $60\%$ of the total epochs.

    The inversion region is set to $[-0.5,0.5]~\text{m} \times [-0.5,0.5]~\text{m}$, discretized into a coarse $64 \times 64$ grid. To quantitatively evaluate the inversion performance of different algorithms, the peak signal-to-noise ratio (PSNR) is adopted as the metric to measure the numerical reconstruction accuracy, where a higher value indicates better quantitative reconstruction accuracy. The mathematical formula for PSNR is:
    \begin{equation}
        \text{PSNR} = 10 \log_{10} \left( \frac{\text{peakval}^2}{\text{MSE}} \right) \quad \text{(dB)}
    \end{equation}
    where ``peakval'' denotes the maximum value of the true relative permittivity or conductivity, and ``MSE'' denotes the mean squared error between the estimate and its reference ground truth.

\subsubsection{Dielectric Targets of Different Contrasts}

    \begin{figure}[!t]
        \hspace*{\fill}%
        \subfloat[]{\includegraphics[width=0.4\linewidth]{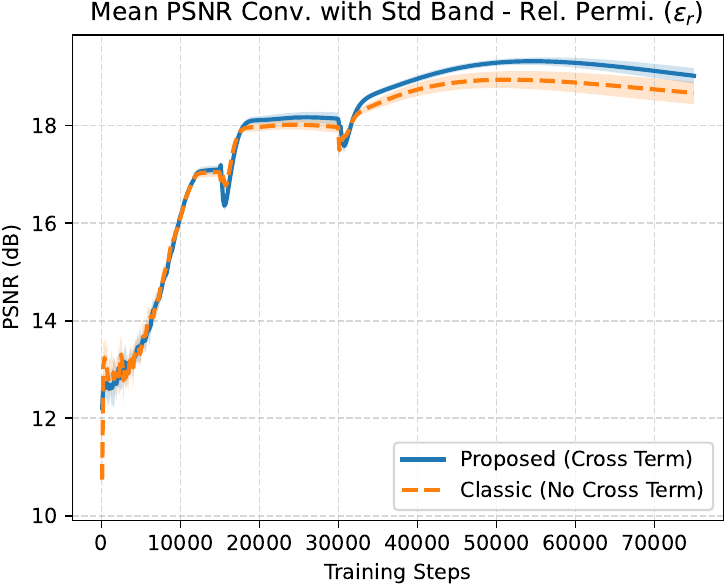}}\hfill
        \subfloat[]{\includegraphics[width=0.4\linewidth]{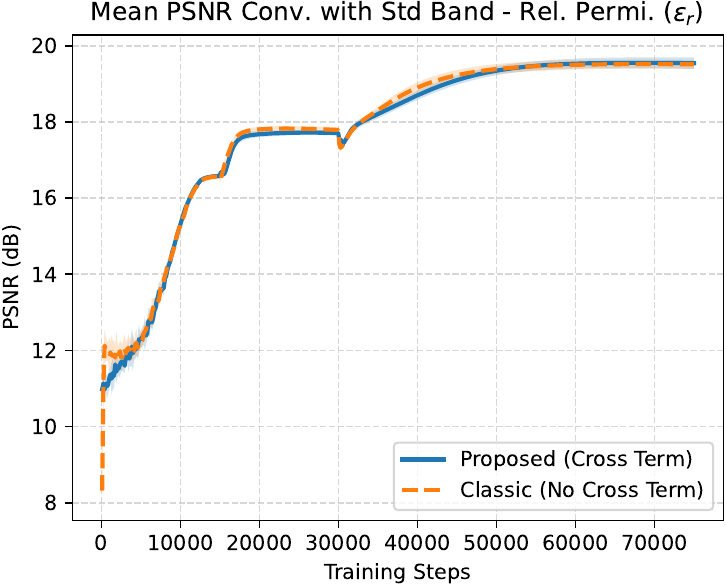}}\hspace*{\fill}

        \hspace*{\fill}%
        \subfloat[]{\includegraphics[width=0.4\linewidth]{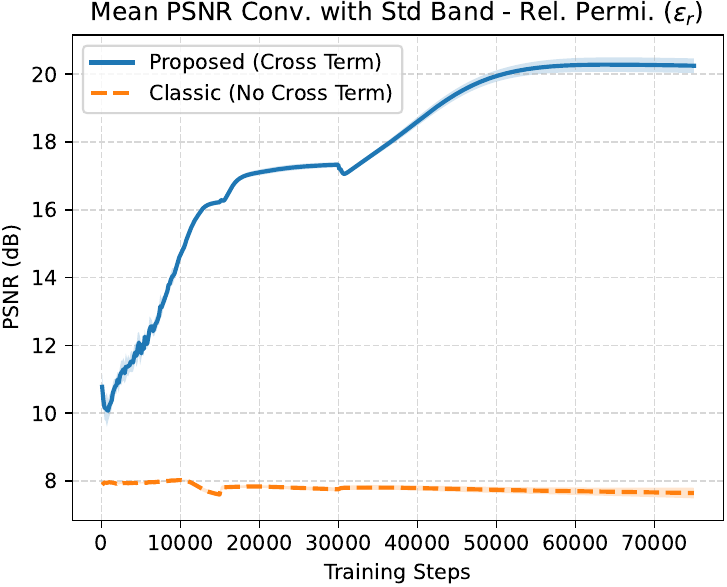}}\hfill
        \subfloat[]{\includegraphics[width=0.4\linewidth]{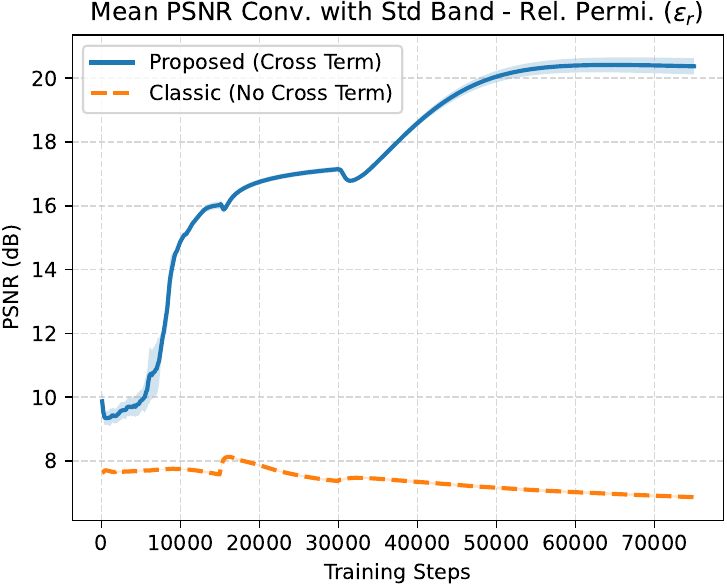}}\hspace*{\fill}

        \hspace*{\fill}%
        \subfloat[]{\includegraphics[width=0.4\linewidth]{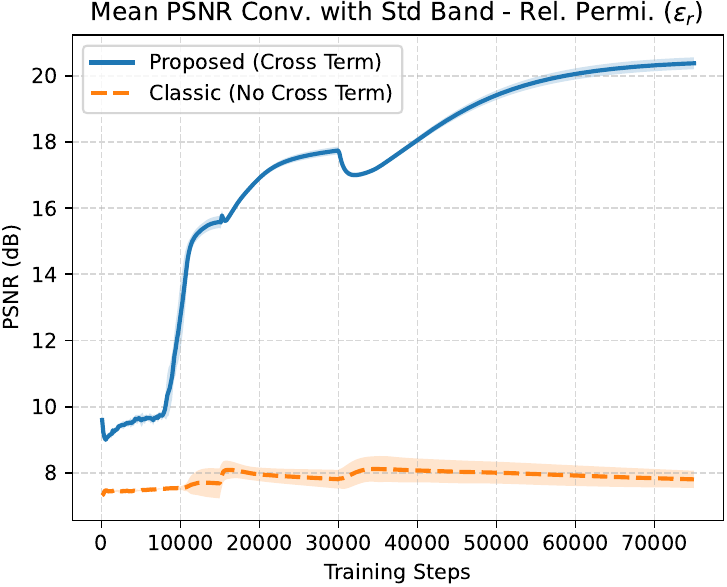}}\hfill
        \subfloat[]{\includegraphics[width=0.4\linewidth]{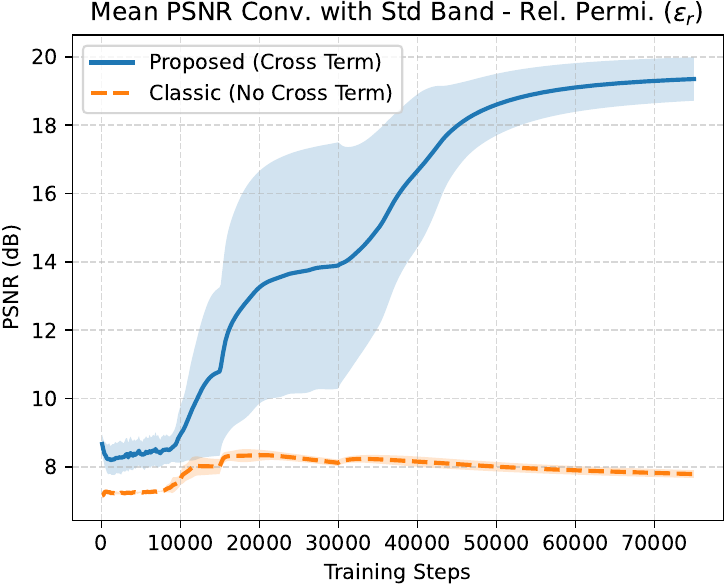}}\hspace*{\fill}
        
        \caption{Comparison of mean PSNR convergence with standard deviation band for CC-PINN and classical cost function PINN using the frequency-hopping strategy to invert ``Austria'' dielectric targets of different contrasts. The relative permittivity of the dielectric targets is $\varepsilon_\text{r}=$4 (a), $\varepsilon_\text{r}=$4.5 (b), $\varepsilon_\text{r}=$5 (c), $\varepsilon_\text{r}=$5.5 (d), $\varepsilon_\text{r}=$6 (e), and $\varepsilon_\text{r}=$6.5 (f). SNR$=20$ dB.}
        \label{fig:mean_psnr_curves_Diel}   
    \end{figure}

    \begin{figure}[!t]
        \hspace*{\fill}%
        \subfloat[]{\includegraphics[width=0.4\linewidth]{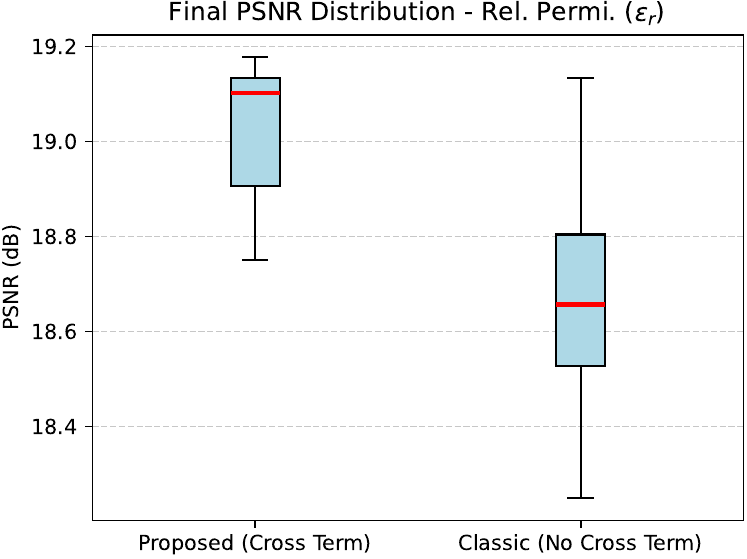}}\hfill
        \subfloat[]{\includegraphics[width=0.4\linewidth]{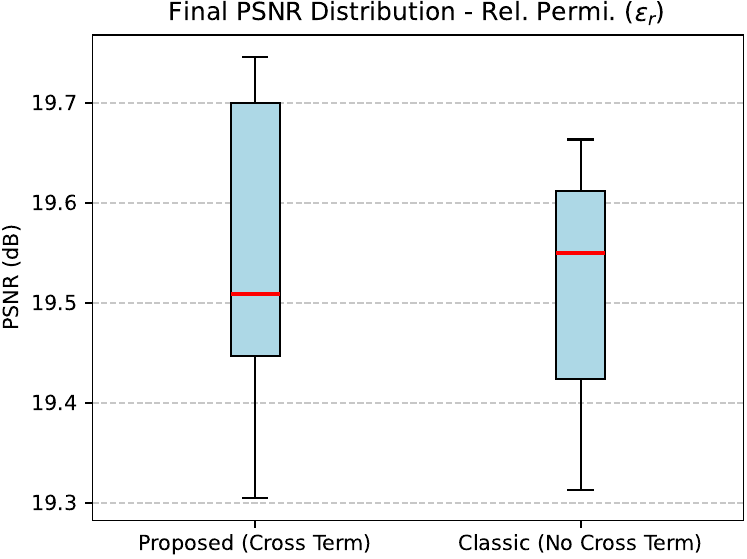}}\hspace*{\fill}

        \hspace*{\fill}%
        \subfloat[]{\includegraphics[width=0.4\linewidth]{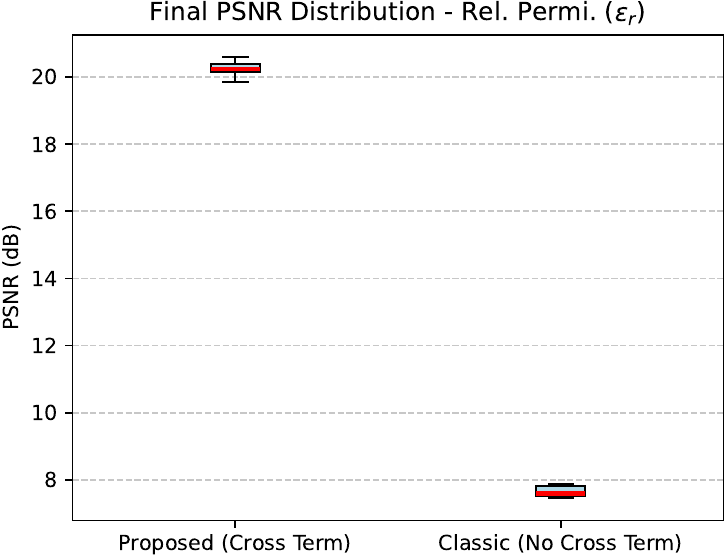}}\hfill
        \subfloat[]{\includegraphics[width=0.4\linewidth]{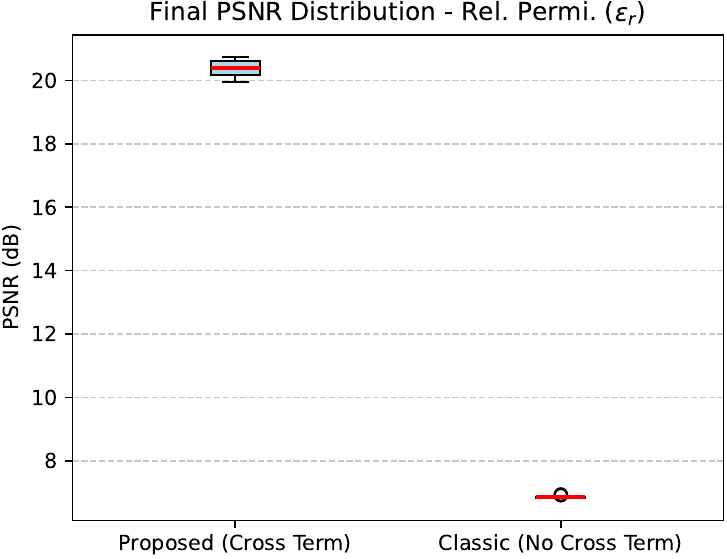}}\hspace*{\fill}

        \hspace*{\fill}%
        \subfloat[]{\includegraphics[width=0.4\linewidth]{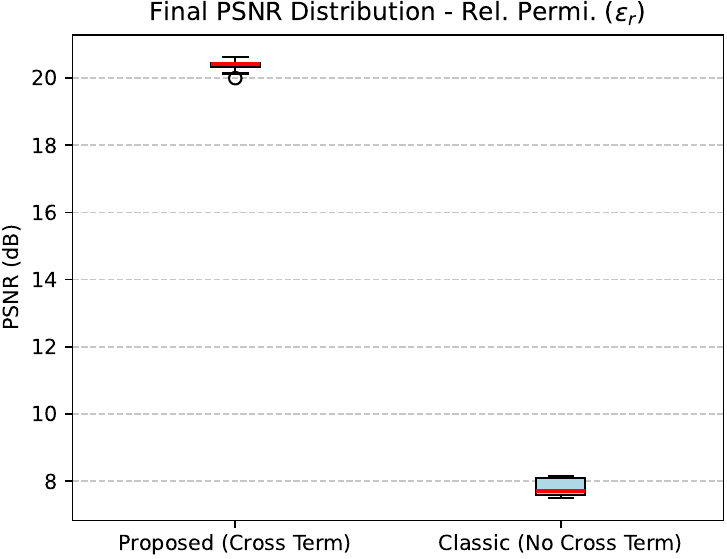}}\hfill
        \subfloat[]{\includegraphics[width=0.4\linewidth]{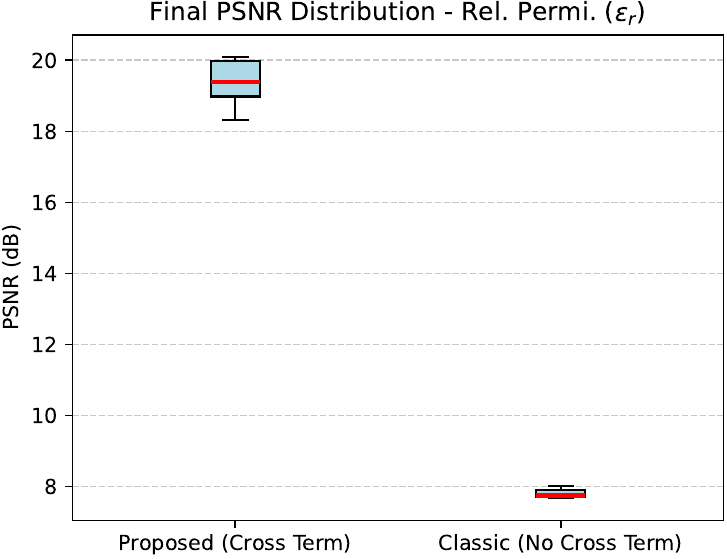}}\hspace*{\fill}
        \caption{Boxplots of final PSNR for CC-PINN and classical cost function PINN using the frequency-hopping strategy to invert ``Austria'' dielectric targets of different contrasts. The relative permittivity of the dielectric targets is $\varepsilon_\text{r}=$4 (a), $\varepsilon_\text{r}=$4.5 (b), $\varepsilon_\text{r}=$5 (c), $\varepsilon_\text{r}=$5.5 (d), $\varepsilon_\text{r}=$6 (e), and $\varepsilon_\text{r}=$6.5 (f). SNR$=20$ dB.}
        \label{fig:boxplot_final_psnr_Diel}   
    \end{figure}

    \begin{figure}[!t]
        \hspace*{\fill}%
        \subfloat[]{\includegraphics[width=0.25\linewidth]{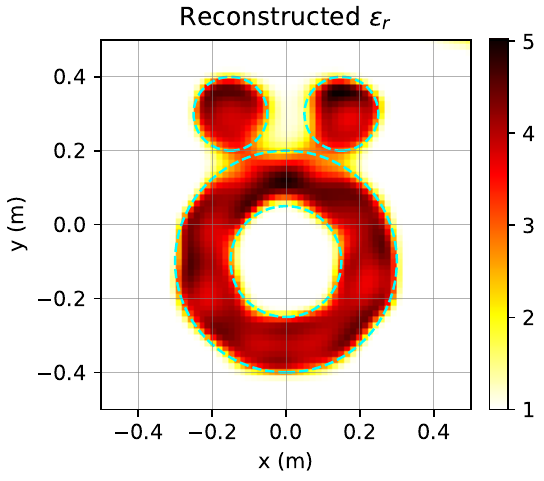}}\hfill
        \subfloat[]{\includegraphics[width=0.25\linewidth]{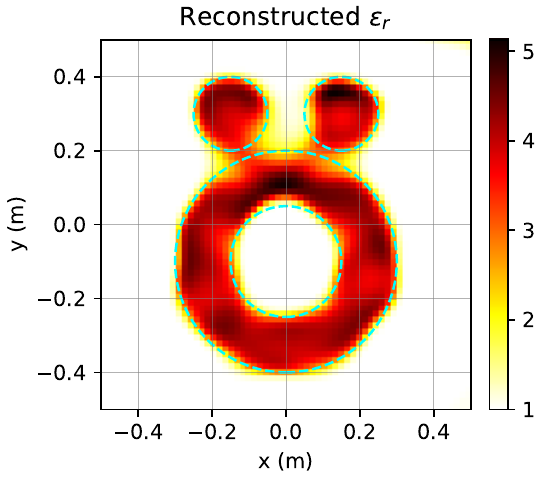}}\hfill
        \subfloat[]{\includegraphics[width=0.25\linewidth]{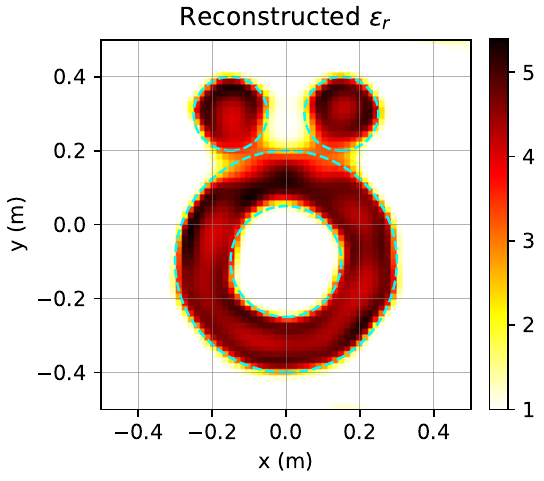}}\hfill
        \subfloat[]{\includegraphics[width=0.25\linewidth]{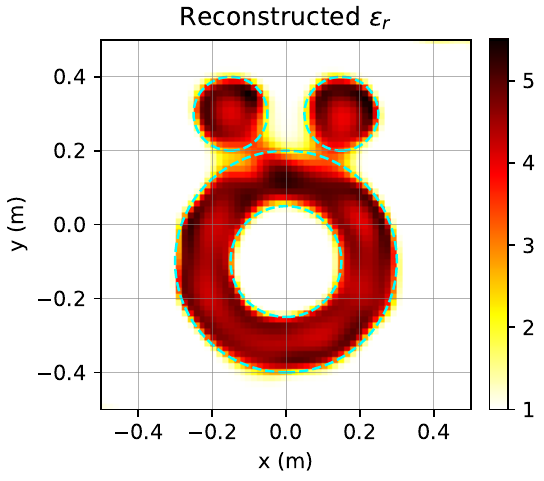}}\hspace*{\fill}

        \hspace*{\fill}%
        \subfloat[]{\includegraphics[width=0.25\linewidth]{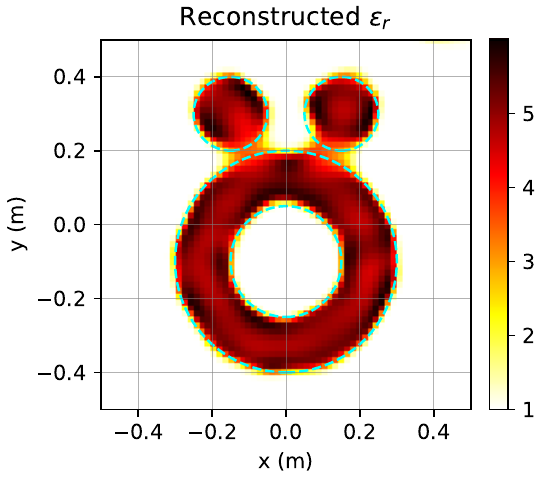}}\hfill
        \subfloat[]{\includegraphics[width=0.25\linewidth]{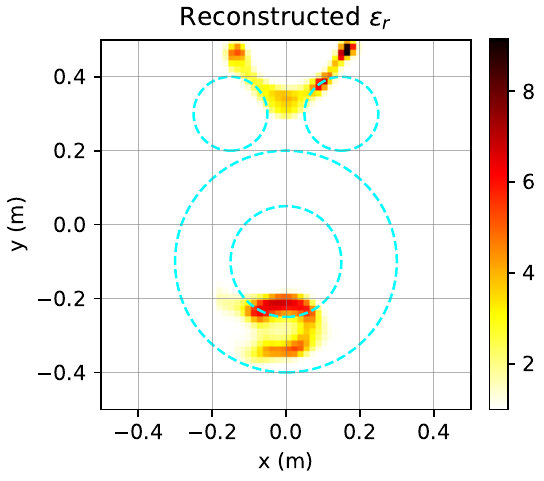}}\hfill
        \subfloat[]{\includegraphics[width=0.25\linewidth]{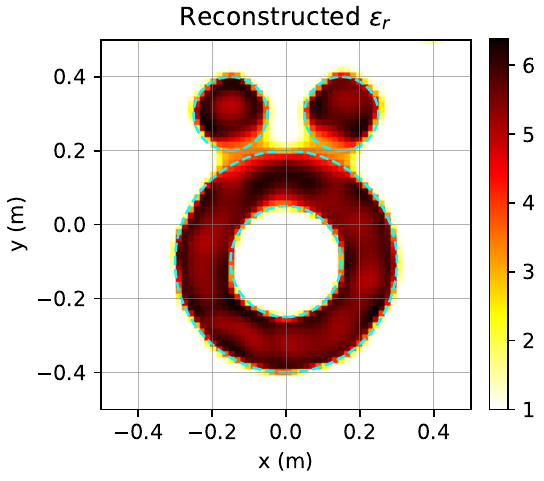}}\hfill
        \subfloat[]{\includegraphics[width=0.25\linewidth]{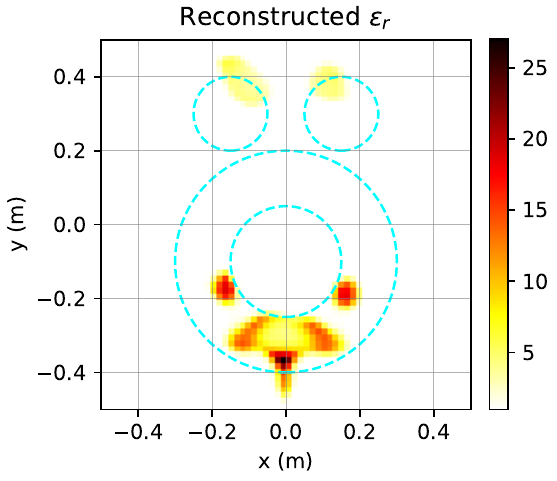}}\hspace*{\fill}

        \hspace*{\fill}%
        \subfloat[]{\includegraphics[width=0.25\linewidth]{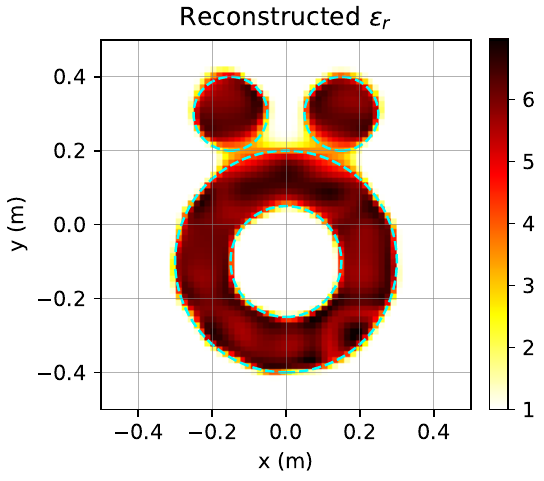}}\hfill
        \subfloat[]{\includegraphics[width=0.25\linewidth]{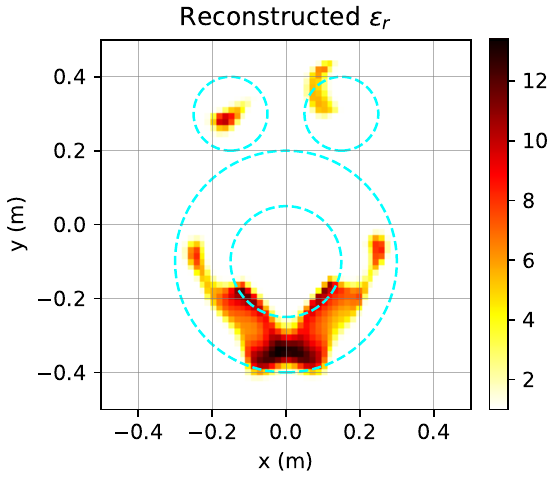}}\hfill
        \subfloat[]{\includegraphics[width=0.25\linewidth]{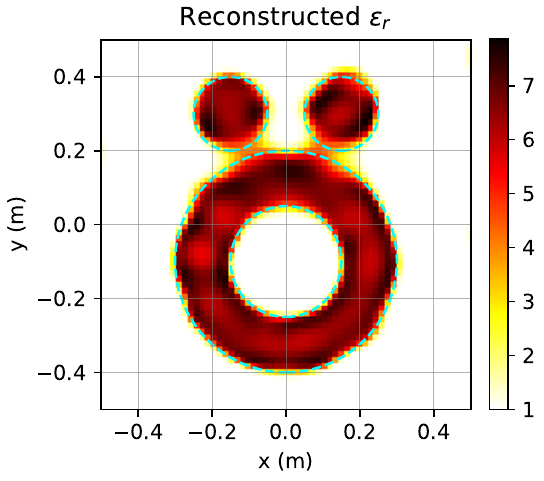}}\hfill
        \subfloat[]{\includegraphics[width=0.25\linewidth]{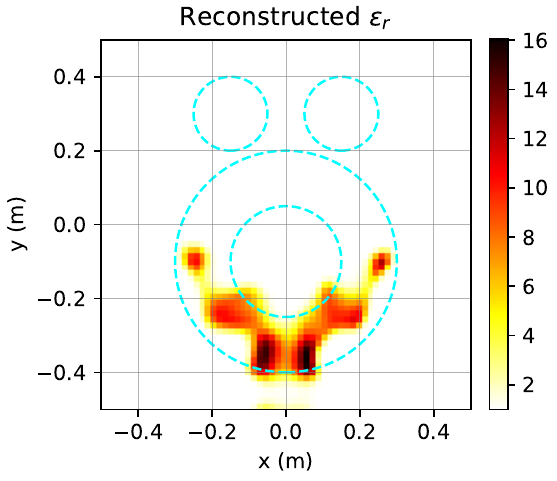}}\hspace*{\fill}
        \caption{Final reconstructed images for CC-PINN and classical cost function PINN using the frequency-hopping strategy to invert ``Austria'' dielectric targets of different contrasts. The relative permittivity of the dielectric targets is $\varepsilon_\text{r}=$4 (a, b), $\varepsilon_\text{r}=$4.5 (c, d), $\varepsilon_\text{r}=$5 (e, f), $\varepsilon_\text{r}=$5.5 (g, h), $\varepsilon_\text{r}=$6 (i, j), and $\varepsilon_\text{r}=$6.5 (k, l). SNR$=20$ dB. Left: CC-PINN; Right: classical cost function PINN. The results shown are selected from the median of the final PSNR of the 11 independent runs. Dashed lines mark the true contours of the target.}
        \label{fig:eps_Diel}   
    \end{figure}

    First, using the frequency-hopping strategy introduced earlier, the relative permittivity of the three cylindrical targets in the ``Austria'' profile is set to $\varepsilon_\text{r}=$4, $\varepsilon_\text{r}=$4.5, $\varepsilon_\text{r}=$5, $\varepsilon_\text{r}=$5.5, $\varepsilon_\text{r}=$6, and $\varepsilon_\text{r}=$6.5, with the SNR fixed at 20 dB. Using the weight-normalized Fourier feature network architecture and initialization parameter settings introduced previously, the traditional cost function and the cross-correlated cost function are respectively employed to perform inversion tests on the above ``Austria'' targets with different relative permittivities.

    To evaluate the final performance and convergence stability of different algorithms over 11 independent runs, this paper employs mean curves and boxplots for analysis. The PSNR is recorded every 100 epochs during training, forming a time series. The mean curve $\mu_t$ reflects the evolution of average performance, and the standard deviation band $\mu_t \pm \sigma_t$ indicates the degree of dispersion among runs; a narrower standard deviation band implies the algorithm is less sensitive to random initialization. The boxplot summarizes the final PSNR samples of each algorithm at the end of the 11 runs; the rectangular box extends from $Q_1$ to $Q_3$, with the internal red line marking the median; the upper and lower whiskers extend to the minimum/maximum values (if there are no outliers). This plot provides an intuitive comparison of the final PSNR level, fluctuation range, and outlier tendency of different algorithms.

    Fig.~\ref{fig:mean_psnr_curves_Diel} and Fig.~\ref{fig:boxplot_final_psnr_Diel} respectively present the mean PSNR curves and boxplots of the final PSNR values for CC-PINN and the PINN using the traditional cost function in this set of tests (six dielectric targets with different permittivities). From the results in the two figures, it can be seen that as the relative permittivity gradually increases, the PINN with the traditional cost function fails to correctly converge to the ground truth starting from $\varepsilon_\text{r}=$5, whereas CC-PINN can still effectively converge to the ground truth up to $\varepsilon_\text{r}=$6.5. This set of results proves that CC-PINN exhibits significantly better robustness than the PINN with the traditional cost function, and for the inversion of the two targets with $\varepsilon_\text{r}=$4 and $\varepsilon_\text{r}=$4.5, the inversion accuracy of the two methods is not much different. To present the inversion results more intuitively, Fig.~\ref{fig:eps_Diel} shows the inversion results of both methods for the six different dielectric targets in this set of experiments. The results in the figure all correspond to the median of the final PSNR from their respective 11 independent runs.

\subsubsection{Performance Analysis under Different Signal-to-Noise Ratios}

    \begin{figure}[!t]
        \hspace*{\fill}%
        \subfloat[]{\includegraphics[width=0.4\linewidth]{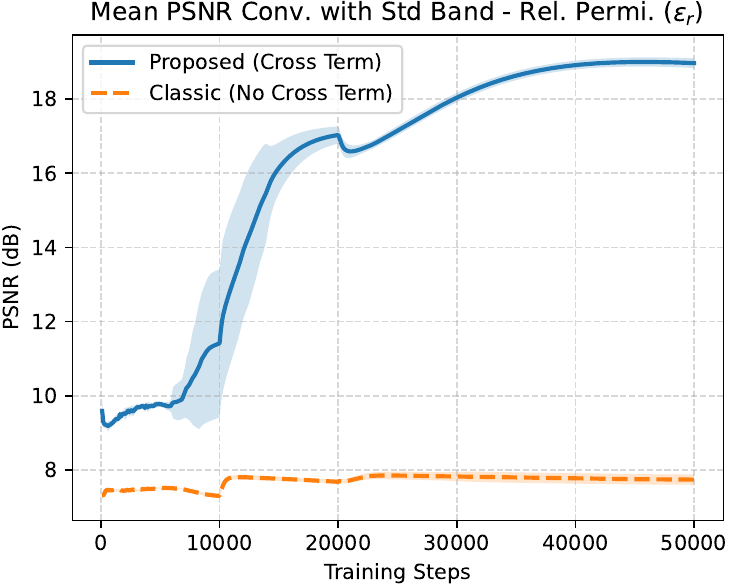}}\hfill
        \subfloat[]{\includegraphics[width=0.4\linewidth]{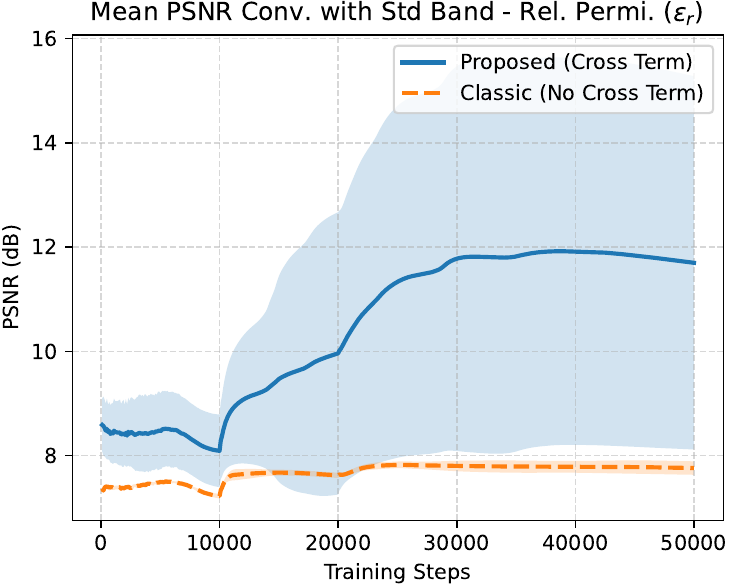}}\hspace*{\fill}

        \hspace*{\fill}%
        \subfloat[]{\includegraphics[width=0.4\linewidth]{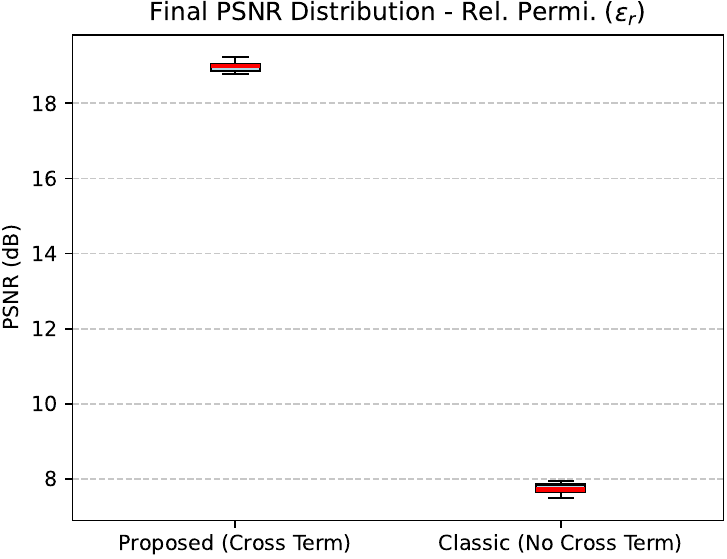}}\hfill
        \subfloat[]{\includegraphics[width=0.4\linewidth]{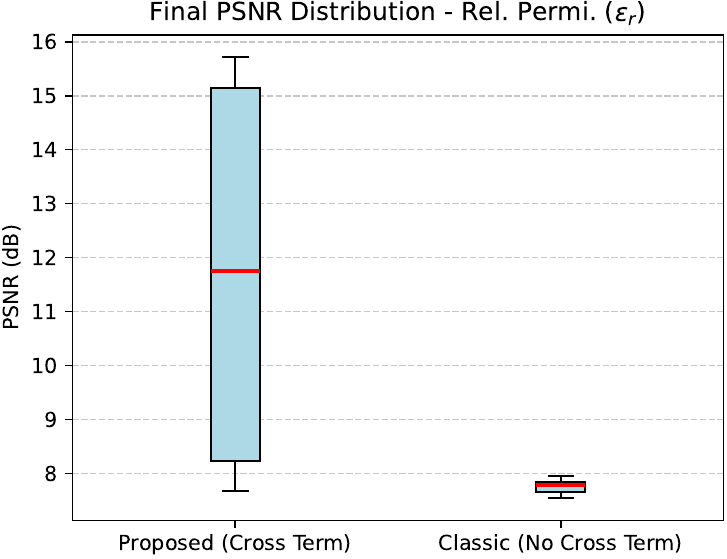}}\hspace*{\fill}
        \caption{Comparison of mean PSNR convergence with standard deviation band and boxplots for CC-PINN and classical cost function PINN using the frequency-hopping strategy to invert the same ``Austria'' dielectric target ($\varepsilon_\text{r}=$6). SNR$=10$ dB (a, c); SNR$=0$ dB (b, d).}
        \label{fig:Diel-6_lowSNR}   
    \end{figure}

    \begin{figure}[!t]
        \hspace*{\fill}%
        \subfloat[]{\includegraphics[height=0.30\linewidth]{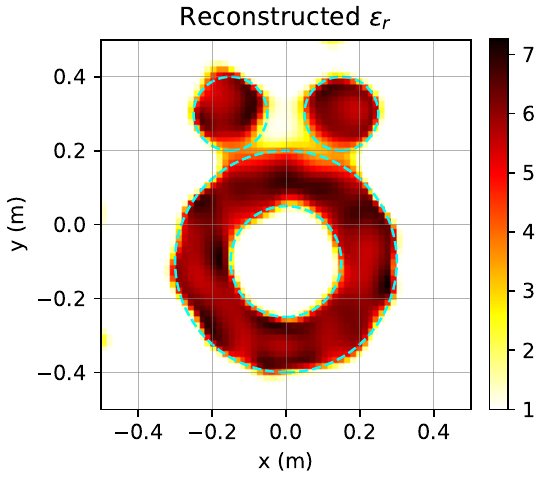}}\hfill
        \subfloat[]{\includegraphics[height=0.30\linewidth]{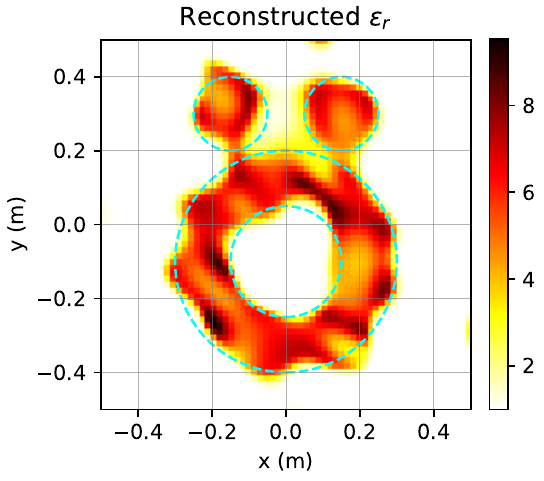}}\hspace*{\fill}

        \hspace*{\fill}%
        \subfloat[]{\includegraphics[height=0.30\linewidth]{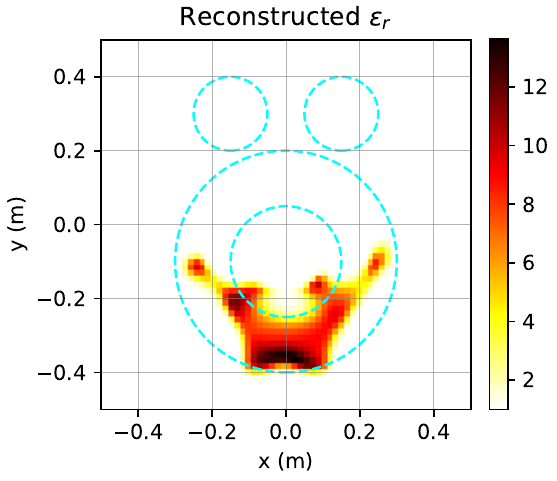}}\hfill
        \subfloat[]{\includegraphics[height=0.30\linewidth]{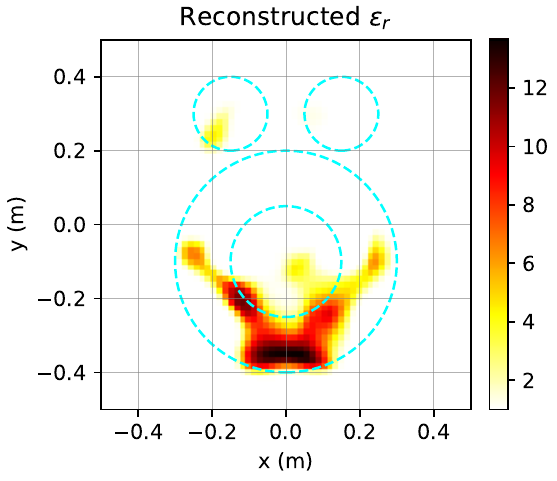}}\hspace*{\fill}
        \caption{Final reconstructed images for CC-PINN and classical cost function PINN using the frequency-hopping strategy to invert the same ``Austria'' dielectric target ($\varepsilon_\text{r}=$6). SNR$=10$ dB (a, c); SNR$=0$ dB (b, d). Top: CC-PINN; Bottom: classical cost function PINN. The results shown are selected from the median of the final PSNR of the 11 independent runs. Dashed lines mark the true contours of the target.}
        \label{fig:Diel-6_10dB_recon}   
    \end{figure}

    In this set of experiments, we select the dielectric target with $\varepsilon_\text{r}=$6 from the previous set and further test the robustness and inversion accuracy of the algorithm under different signal-to-noise ratios. Fig.~\ref{fig:Diel-6_lowSNR} presents the mean PSNR curves and boxplots of CC-PINN and the PINN with the traditional cost function at SNR$=10$ dB and SNR$=0$ dB. From the figure, it can be observed that when the SNR drops to 10 dB, the PSNR curves of CC-PINN begin to exhibit noticeable fluctuations but still converge to the ground truth. When the SNR drops to 0 dB, the PSNR curves of CC-PINN show a wider range of violent fluctuations, and the algorithm's robustness begins to decline. On the other hand, the PINN with the traditional cost function completely fails to converge correctly throughout. Fig.~\ref{fig:Diel-6_10dB_recon} more intuitively presents the inversion results of this set of experiments, where the results all correspond to the median of the final PSNR from their respective 11 independent runs.

\subsubsection{Simultaneous Multi-Frequency Processing Strategy}

    \begin{figure}[!t]
        \hspace*{\fill}%
        \subfloat[]{\includegraphics[width=0.4\linewidth]{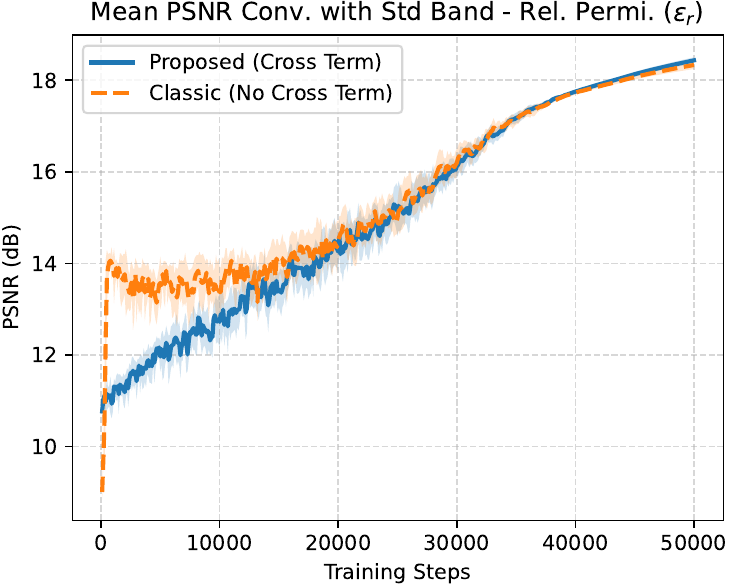}}\hfill
        \subfloat[]{\includegraphics[width=0.4\linewidth]{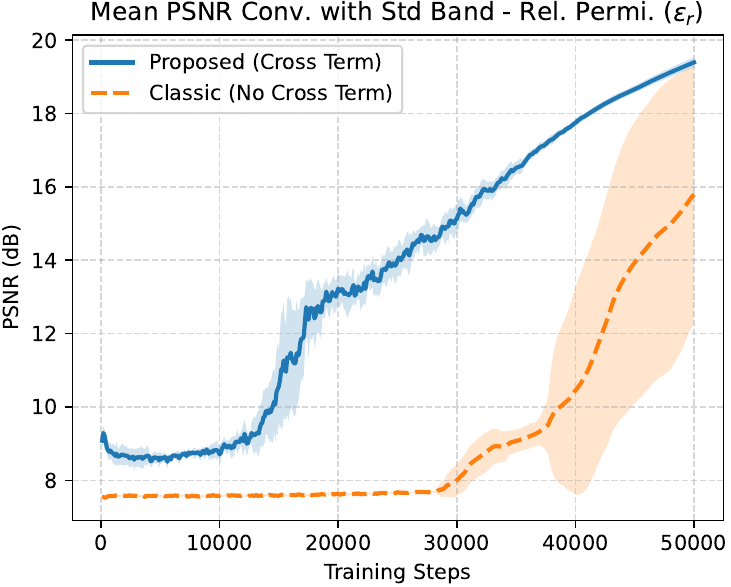}}\hspace*{\fill}
        
        \hspace*{\fill}%
        \subfloat[]{\includegraphics[width=0.4\linewidth]{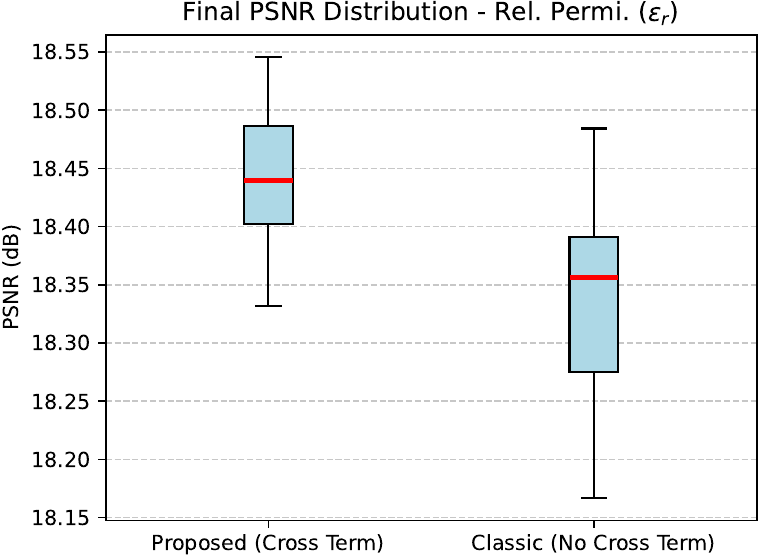}}\hfill
        \subfloat[]{\includegraphics[width=0.4\linewidth]{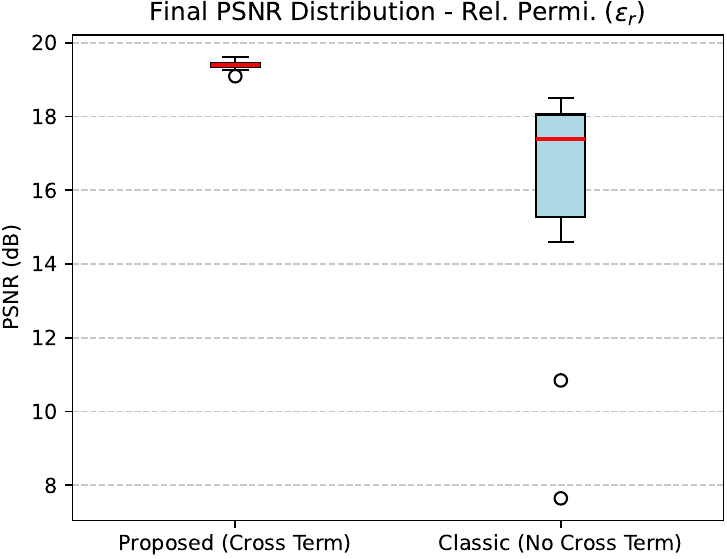}}\hspace*{\fill}
        
        \caption{Comparison of mean PSNR convergence with standard deviation band for CC-PINN and classical cost function PINN using the simultaneous multi-frequency processing strategy to invert ``Austria'' dielectric targets of different contrasts. Target relative permittivity is $\varepsilon_\text{r}=$4 (a, c) and $\varepsilon_\text{r}=$5 (b, d). SNR$=20$ dB.}
        \label{fig:Diel-Simultaneous}   
    \end{figure}

    In this set of experiments, the simultaneous multi-frequency processing strategy is adopted to further test the algorithm performance, with the test targets being the dielectric targets from the first set of experiments, with relative permittivities of $\varepsilon_\text{r}=$4 and $\varepsilon_\text{r}=$5. Fig.~\ref{fig:Diel-Simultaneous} shows the mean PSNR curves and boxplots for the two targets under the simultaneous multi-frequency processing strategy. Comparing these results with those in Fig.~\ref{fig:mean_psnr_curves_Diel} and Fig.~\ref{fig:boxplot_final_psnr_Diel}, it can be found that in these several sets of simulation experiments, the convergence performance and inversion accuracy of simultaneous multi-frequency processing are significantly worse than those of the frequency-hopping strategy. The mean PSNR curves exhibit pronounced fluctuations almost throughout the entire process.

    The authors provide the following explanation for this phenomenon: In the synthetic data, the scatterers, incident field, and boundary conditions fully conform to the Helmholtz equation. Frequency hopping starts from the lowest frequency, obtaining a smooth but topologically correct coarse-resolution solution, which still serves as a good initial point under the high-frequency equation gradients, thus enabling progressive refinement and avoiding local minima of strong nonlinearity at high frequencies. This is a well-established basis for the success of frequency hopping in classical inverse scattering theory. When we adopt the PINN architecture to process multi-frequency synthetic data simultaneously, the low-frequency losses tend to adjust large-scale contours, while high-frequency losses are extremely sensitive to interface details. When the network parameters receive gradients from both directions in a single update, they may be mutually orthogonal or even completely opposite, causing the parameter update direction to oscillate violently, with the PSNR fluctuating accordingly. It is worth mentioning that in the subsequent measured data processing, we will observe the exact opposite phenomenon, i.e., when processing measured data, simultaneous multi-frequency processing is instead more likely to converge to the true dielectric parameter distribution. The authors believe this is caused by the differences between measured data and synthetic data, which will be discussed in detail in a later section.

\subsubsection{Lossy Dielectric Targets}

    \begin{figure}[!t]
        \hspace*{\fill}%
        \subfloat[]{\includegraphics[width=0.4\linewidth]{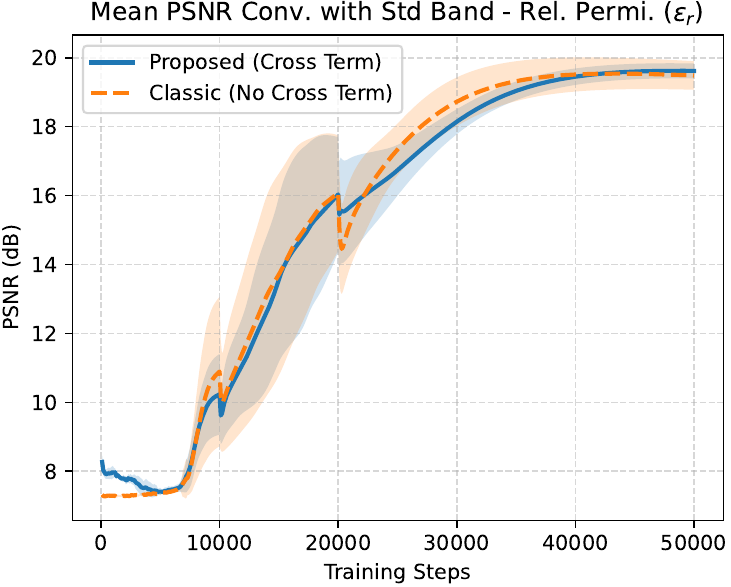}}\hfill
        \subfloat[]{\includegraphics[width=0.4\linewidth]{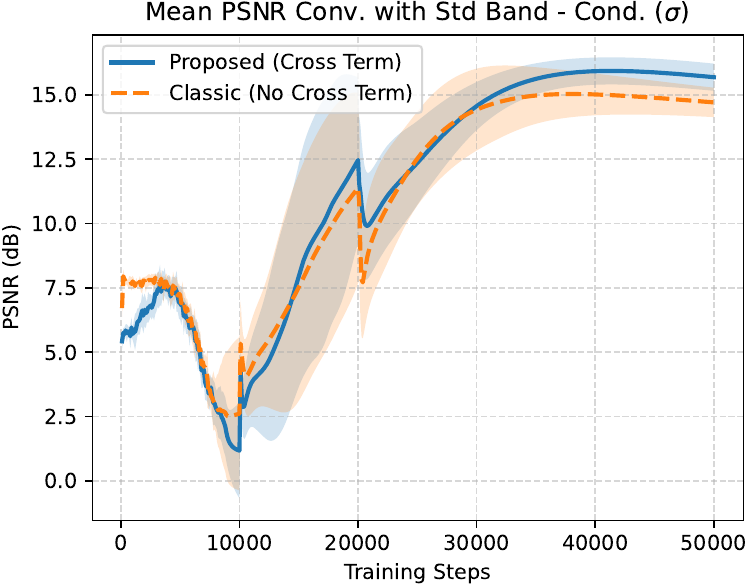}}\hspace*{\fill}

        \hspace*{\fill}%
        \subfloat[]{\includegraphics[width=0.4\linewidth]{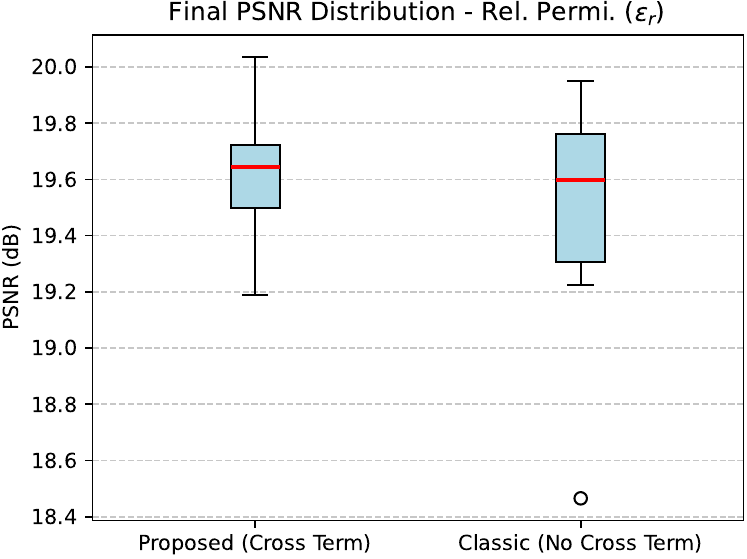}}\hfill
        \subfloat[]{\includegraphics[width=0.4\linewidth]{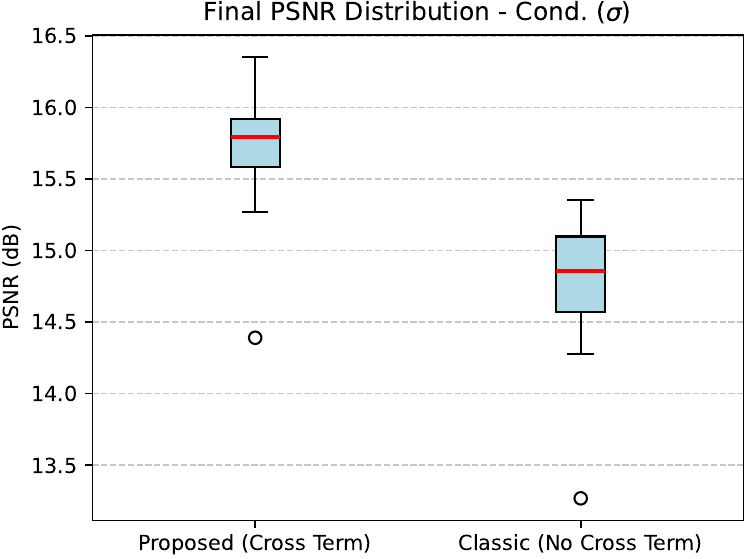}}\hspace*{\fill}
        \caption{Comparison of mean PSNR convergence with standard deviation band and boxplots for CC-PINN and classical cost function PINN using the frequency-hopping strategy to invert ``Austria'' lossy dielectric target 1 ($\varepsilon_\text{r}=6$, $\sigma=0.03$ S/m). SNR$=20$ dB.}
        \label{fig:Lossy-e6s30}   
    \end{figure}

    \begin{figure}[!t]
        \hspace*{\fill}%
        \subfloat[]{\includegraphics[width=0.4\linewidth]{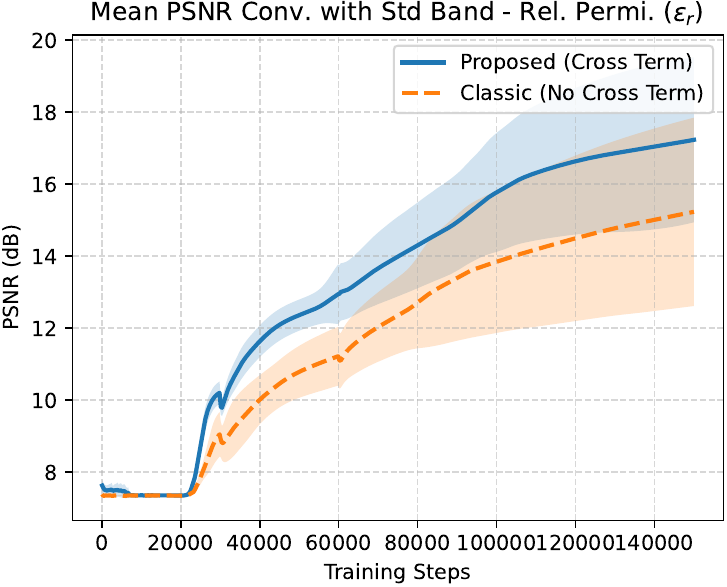}}\hfill
        \subfloat[]{\includegraphics[width=0.4\linewidth]{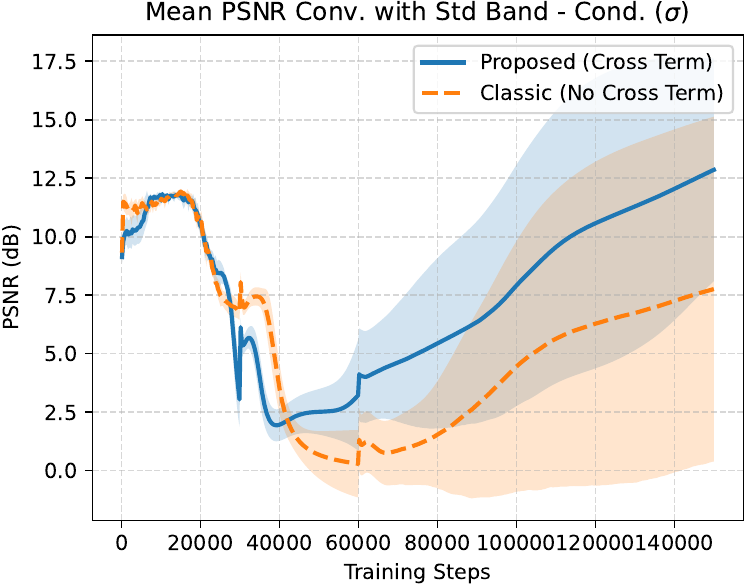}}\hspace*{\fill}
        
        \hspace*{\fill}%
        \subfloat[]{\includegraphics[width=0.4\linewidth]{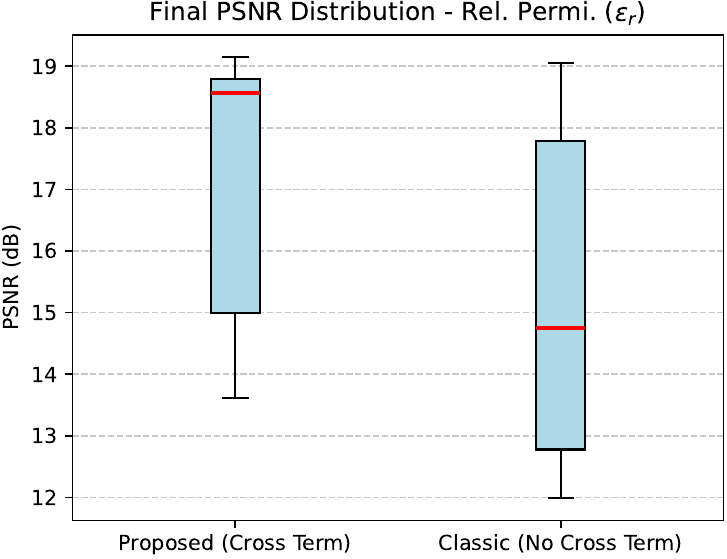}}\hfill
        \subfloat[]{\includegraphics[width=0.4\linewidth]{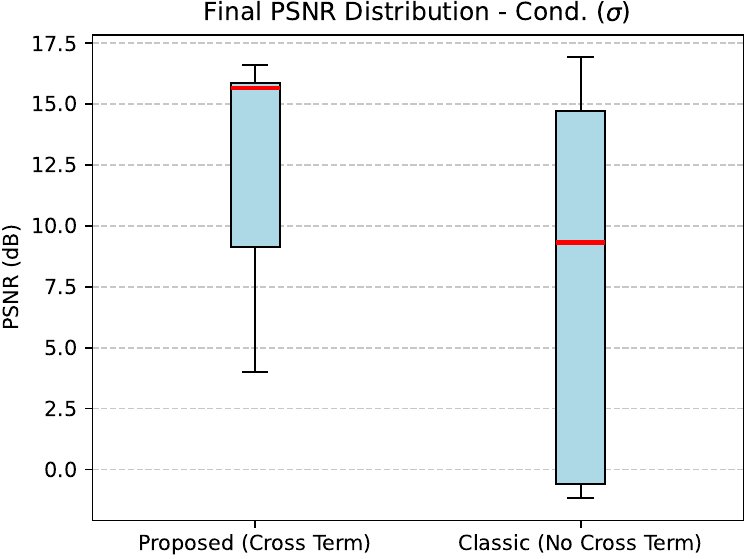}}\hspace*{\fill}
        
        \caption{Comparison of mean PSNR convergence with standard deviation band and boxplots for CC-PINN and classical cost function PINN using the frequency-hopping strategy to invert ``Austria'' lossy dielectric target 2 (small cylinders: $\varepsilon_\text{r}=6$, $\sigma=0.05$ S/m; large ring: $\varepsilon_\text{r}=9$, $\sigma=0.03$ S/m). SNR$=20$ dB.}
        \label{fig:Lossy-e6s50e6s50e9s30}   
    \end{figure}

    \begin{figure}[!t]
        \hspace*{\fill}%
        \subfloat[]{\includegraphics[height=0.21\linewidth]{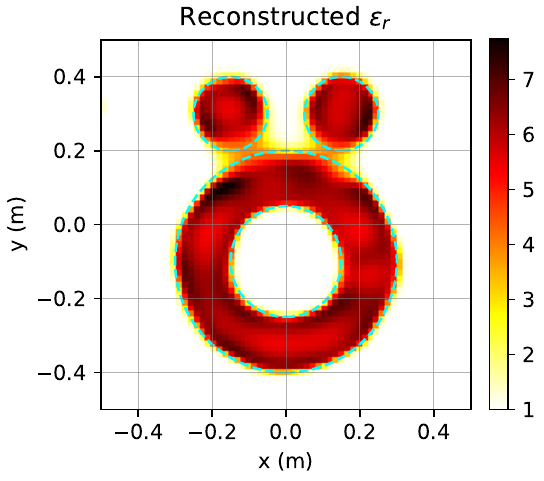}}\hfill
        \subfloat[]{\includegraphics[height=0.21\linewidth]{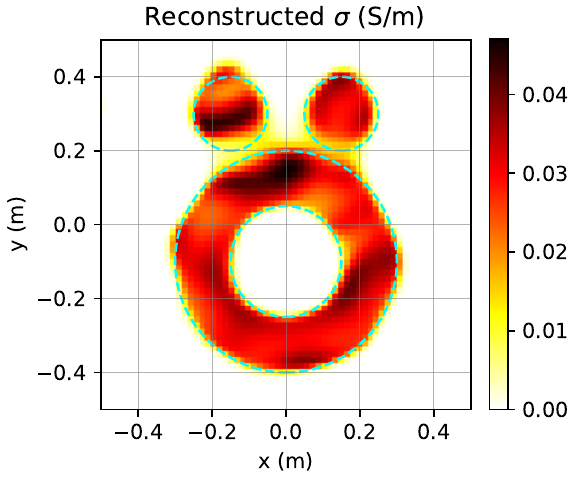}}\hfill
        \subfloat[]{\includegraphics[height=0.21\linewidth]{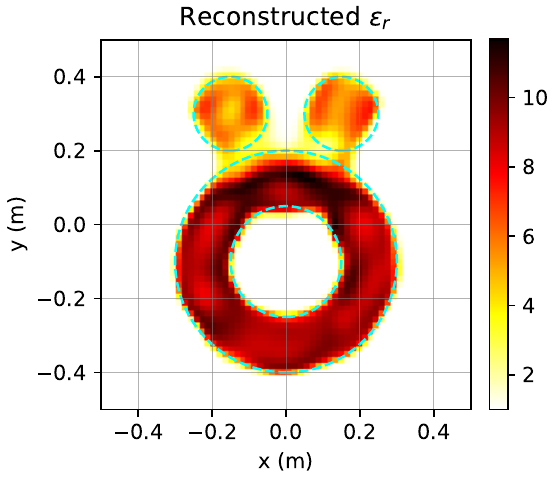}}\hfill
        \subfloat[]{\includegraphics[height=0.21\linewidth]{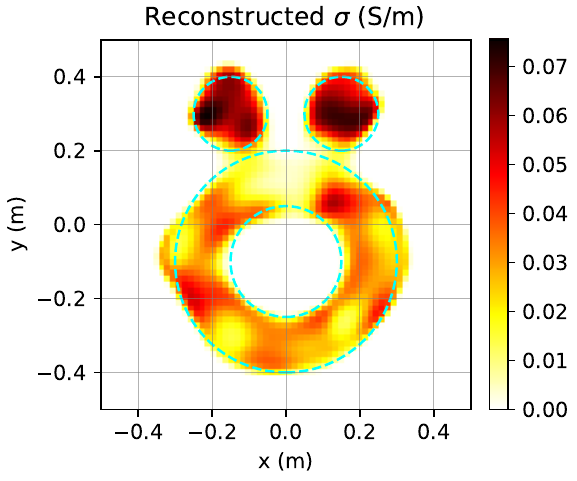}}\hspace*{\fill}

        \hspace*{\fill}%
        \subfloat[]{\includegraphics[height=0.21\linewidth]{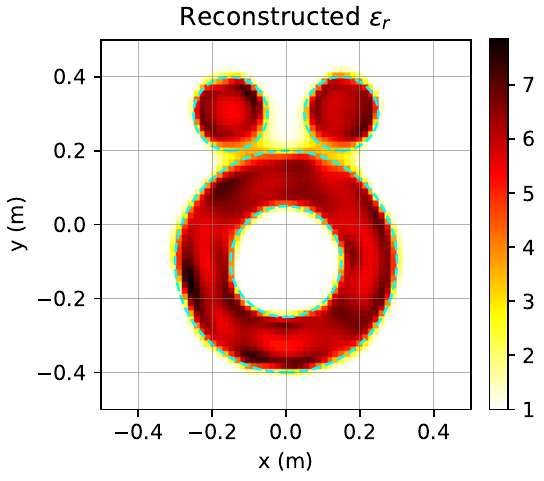}}\hfill
        \subfloat[]{\includegraphics[height=0.21\linewidth]{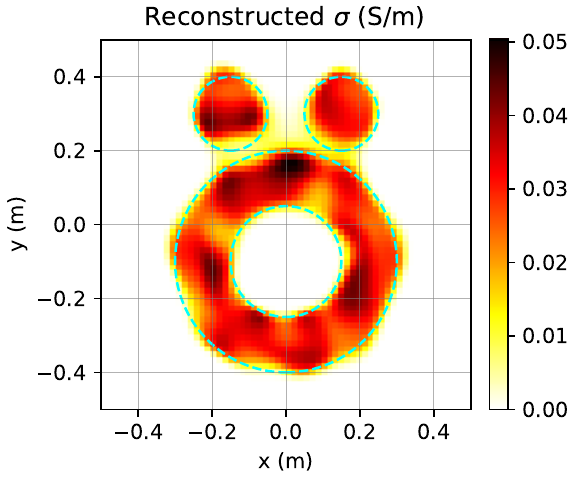}}\hfill
        \subfloat[]{\includegraphics[height=0.21\linewidth]{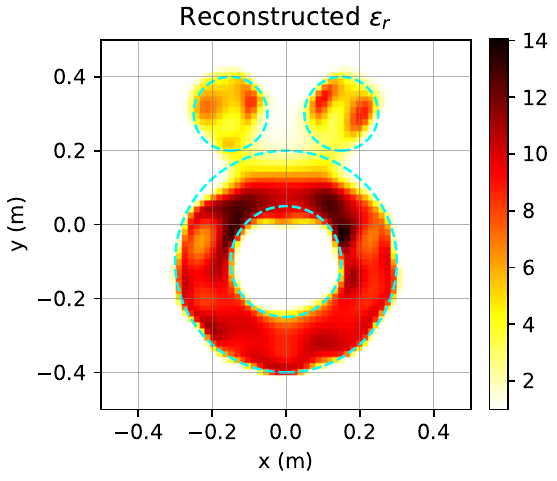}}\hfill
        \subfloat[]{\includegraphics[height=0.21\linewidth]{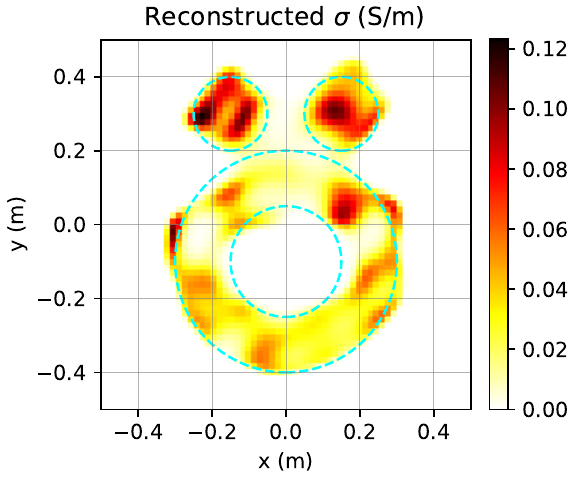}}\hspace*{\fill}
        \caption{Reconstruction results for CC-PINN and classical cost function PINN using the frequency-hopping strategy to invert different lossy ``Austria'' dielectric targets. ``Austria'' lossy target 1: $\varepsilon_\text{r}=6$, $\sigma=0.03$ S/m (a--b, e--f); ``Austria'' lossy target 2: small cylinders $\varepsilon_\text{r}=6$, $\sigma=0.05$ S/m, large ring $\varepsilon_\text{r}=9$, $\sigma=0.03$ S/m (c--d, g--h). SNR$=20$ dB. Left: relative permittivity; Right: conductivity (S/m). Top: CC-PINN; Bottom: classical cost function PINN. The results shown are selected from the median of the final PSNR of the 11 independent runs. Dashed lines mark the true contours of the target.}
        \label{fig:Lossy-recon}   
    \end{figure}

    As the final set of synthetic experiments, we further examine the performance of the two algorithms in inverting lossy dielectric targets. The ``Austria'' target is still used, and two sets of dielectric parameter combinations are prepared from easy to difficult: the first set is uniformly set to $\varepsilon_\text{r}=6,\ \sigma=0.03$ S/m; the second set has the two cylinders set to $\varepsilon_\text{r}=6,\ \sigma=0.05$ S/m, and the ring cylinder set to $\varepsilon_\text{r}=9,\ \sigma=0.03$ S/m. The SNR is set to 20 dB. Fig.~\ref{fig:Lossy-e6s30} and Fig.~\ref{fig:Lossy-e6s50e6s50e9s30} respectively present the mean PSNR curves and boxplots of relative permittivity and conductivity for the two sets of targets. From the results in the figures, it can be seen that for inverting the first set of targets, the inversion accuracy and robustness of the two methods are basically consistent, while for the more difficult second set of targets, CC-PINN demonstrates better inversion accuracy and robustness. Fig.~\ref{fig:Lossy-recon} shows the inversion results of this set of experiments, with the results all corresponding to the median of the final PSNR from their respective 11 independent runs. It should be noted that due to the presence of conductivity in lossy media, the field distribution inside the target attenuates rapidly, which greatly reduces the nonlinearity and ill-posedness of the electromagnetic inverse scattering problem, thereby explaining why inversion algorithms can reconstruct larger values of relative permittivity when inverting lossy media.

\subsection{Validation with Measured Data}

\begin{figure}[!t]
        \hspace*{\fill}%
        \subfloat[]{\includegraphics[width=0.32\linewidth]{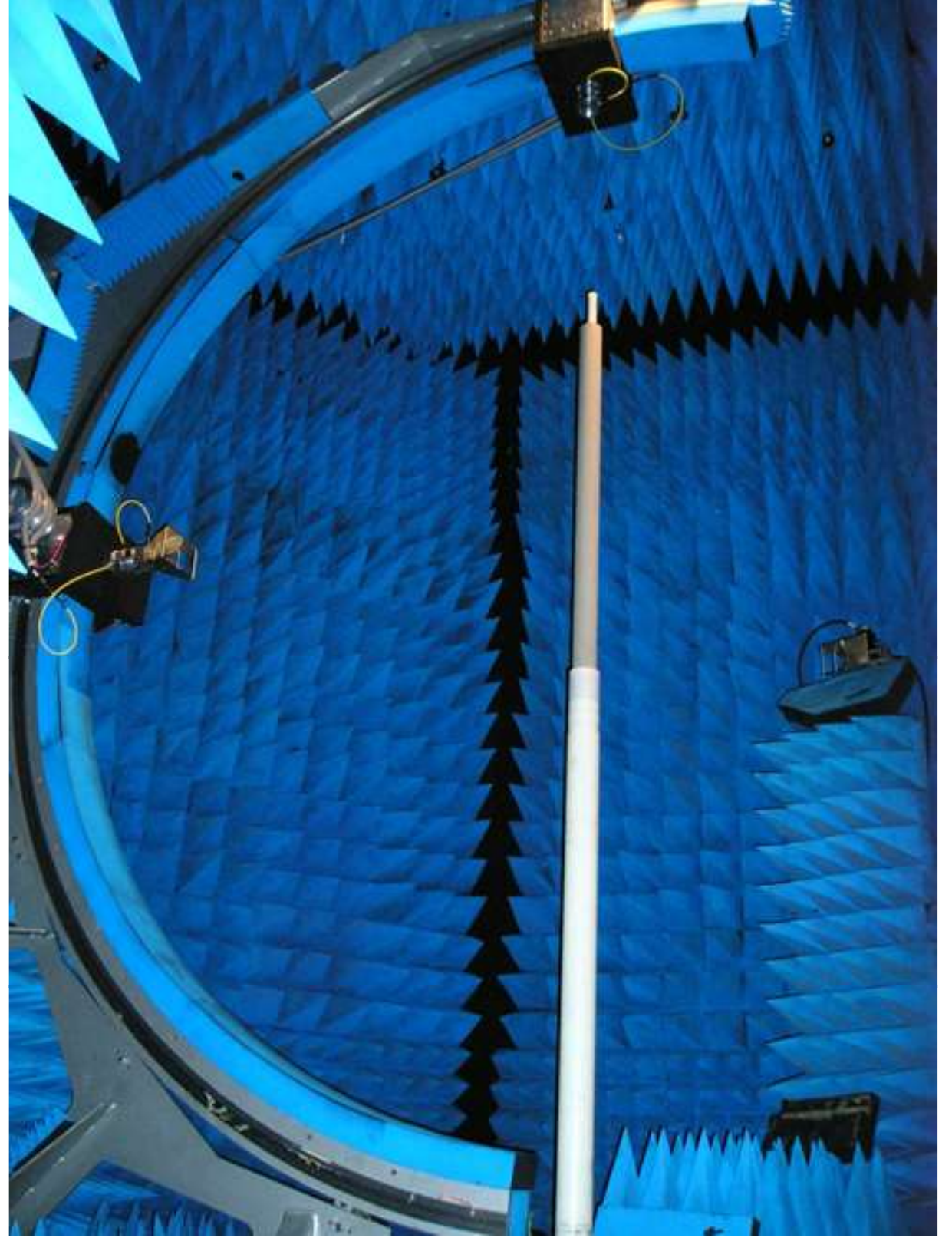}}\hfill
        \subfloat[]{\includegraphics[width=0.50\linewidth]{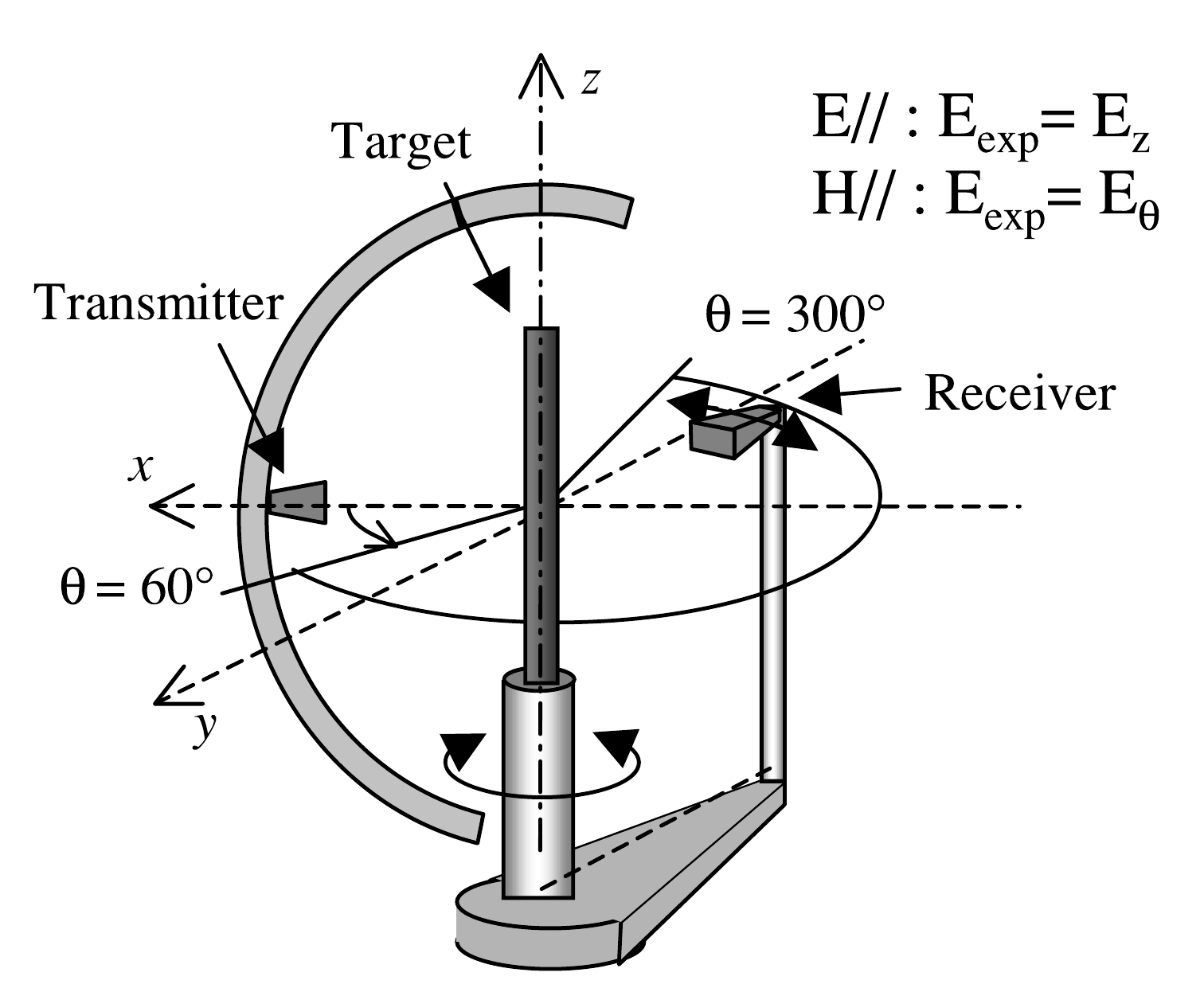}}\hspace*{\fill}
        \caption{Free space scattering measurement facility of the Fresnel data. (a) Photograph of the microwave anechoic chamber experimental setup; (b) Schematic diagram of the geometric relationships in the detection scenario.}
        \label{fig:FresnelScene}   
    \end{figure}

    In this section, experimentally measured data from the Fresnel Institute are used to further validate the proposed method. We use the TM-polarized dataset \textit{FoamTwinDielTM} released in 2005, which contains one large dielectric cylinder ($\varepsilon_r = 1.45\pm 0.15$, diameter 80 mm) and two smaller dielectric cylinders ($\varepsilon_r = 3\pm 0.3$, diameter 31 mm). Fig.~\ref{fig:FresnelScene} shows the measurement configuration. The original experimental setup contains 18 incident angles, with 241 spatial sampling points distributed along an arc of radius 1.67 m, at 9 discrete frequencies between 2 GHz and 10 GHz ($f_1=2$ GHz, $f_2=3$ GHz, $\cdots$, $f_9=10$ GHz), yielding a total of $18\times 241\times 9$ complex scattered field measurements. More information about this two-dimensional Fresnel dataset can be found in~\cite{geffrin2005free}.

    In the subsequent data processing, all 18 equally spaced incident angles will be used, and the angular interval between receiving elements is simplified to 5$^\circ$, resulting in a reduced dataset containing $18\times 49\times 9$ complex measurements. The inversion region is set to [$-$0.1 m, 0.1 m] $\times$ [$-$0.1 m, 0.1 m], with the same number of grid cells as in the simulation processing, discretized into a $64\times 64$ two-dimensional grid. To quantitatively evaluate the inversion accuracy and robustness of the algorithms, the center positions of the cylinders are estimated from good inversion results, and the true relative permittivities of the large cylinder and the two smaller cylinders are determined as $\varepsilon_r = 1.45$, $\varepsilon_r = 3$, and $\varepsilon_r = 3$, based on which the PSNR of the reconstructed images is computed.

    Based on the frequency values, the dataset is divided into a low-frequency dataset ``FoamTwinDielTM\_345'' ($3$ GHz $+$ $4$ GHz $+$ $5$ GHz) and a high-frequency dataset ``FoamTwinDielTM\_678'' ($6$ GHz $+$ $7$ GHz $+$ $8$ GHz). The high-frequency dataset has stronger nonlinearity and ill-posedness, which can effectively increase the inversion difficulty of the same target and is used to further test the robustness of the proposed algorithm. The frequency-hopping strategy described earlier is attempted, with the total number of epochs set to 15000, Stage 1 ($3$ GHz) and Stage 2 ($3$ GHz $+$ $4$ GHz) equally sharing $40\%$ of the total epochs, and Stage 3 ($3$ GHz $+$ $4$ GHz $+$ $5$ GHz) accounting for $60\%$ of the total epochs.

    \begin{figure}[!t]
        \hspace*{\fill}%
        \subfloat[]{\includegraphics[width=0.4\linewidth]{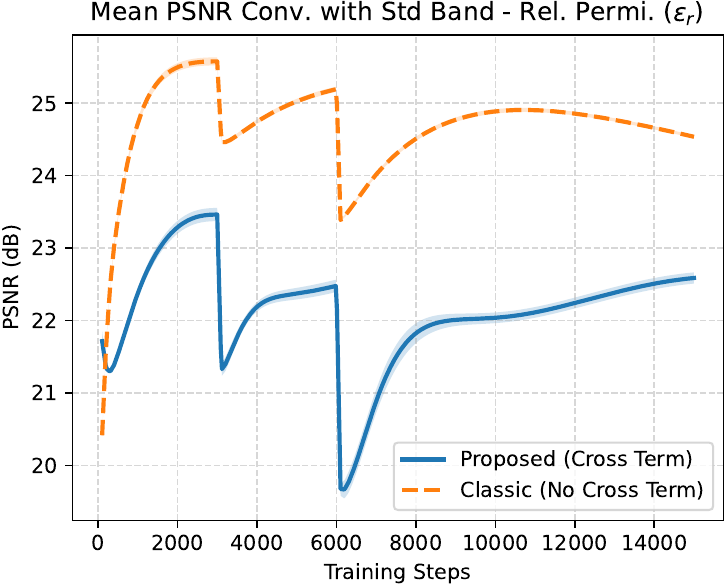}}\hfill
        \subfloat[]{\includegraphics[width=0.4\linewidth]{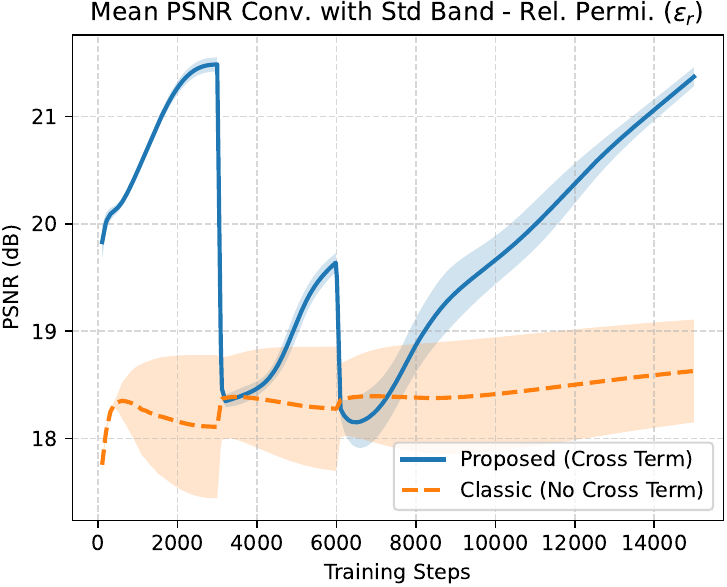}}\hspace*{\fill}
        
        \hspace*{\fill}%
        \subfloat[]{\includegraphics[width=0.4\linewidth]{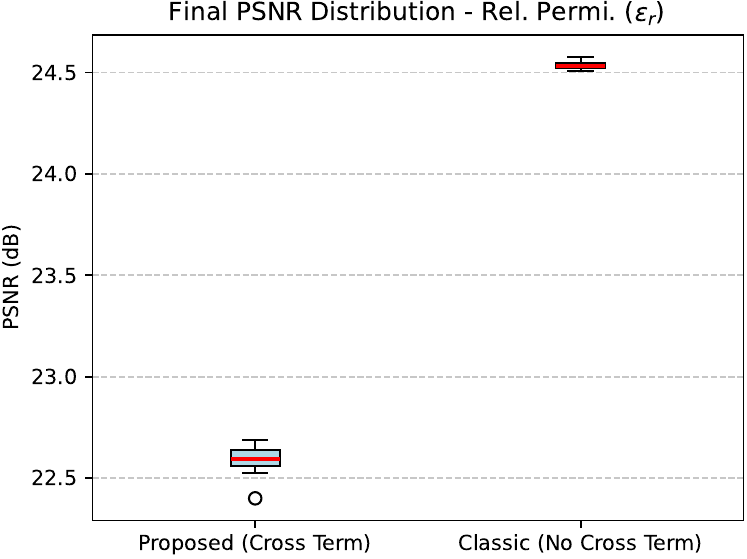}}\hfill
        \subfloat[]{\includegraphics[width=0.4\linewidth]{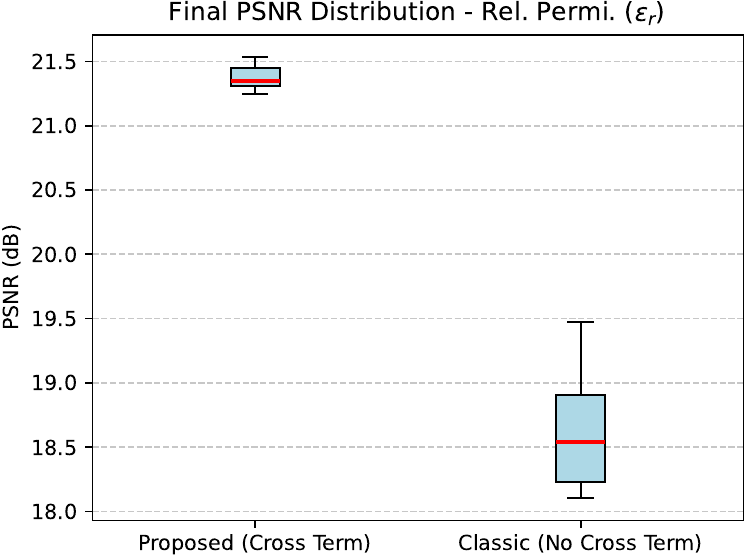}}\hspace*{\fill}
        
        \caption{Comparison of mean PSNR convergence with standard deviation band and boxplots for CC-PINN and classical cost function PINN using the frequency-hopping strategy to invert the Fresnel low-frequency dataset \textit{FoamTwinDielTM\_345} (a, c) and high-frequency dataset \textit{FoamTwinDielTM\_678} (b, d).}
        \label{fig:FoamTwinDielTM_FreqHp}   
    \end{figure}

    \begin{figure}[!t]
        \hspace*{\fill}%
        \subfloat[]{\includegraphics[width=0.4\linewidth]{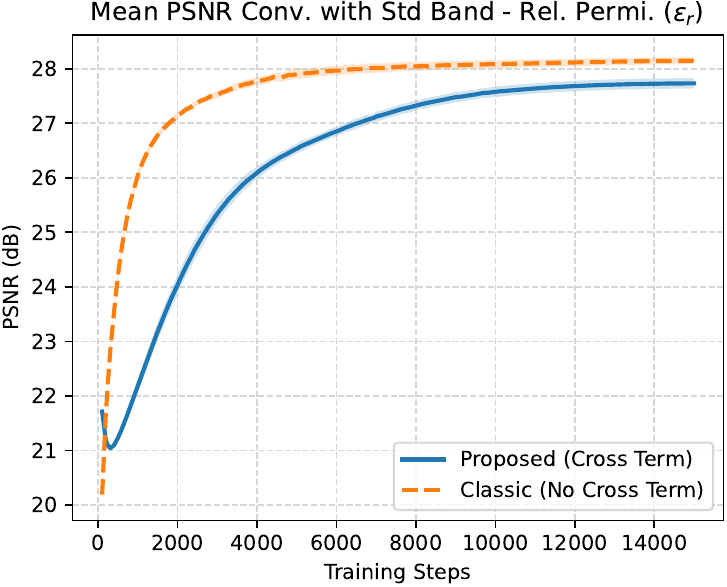}}\hfill
        \subfloat[]{\includegraphics[width=0.4\linewidth]{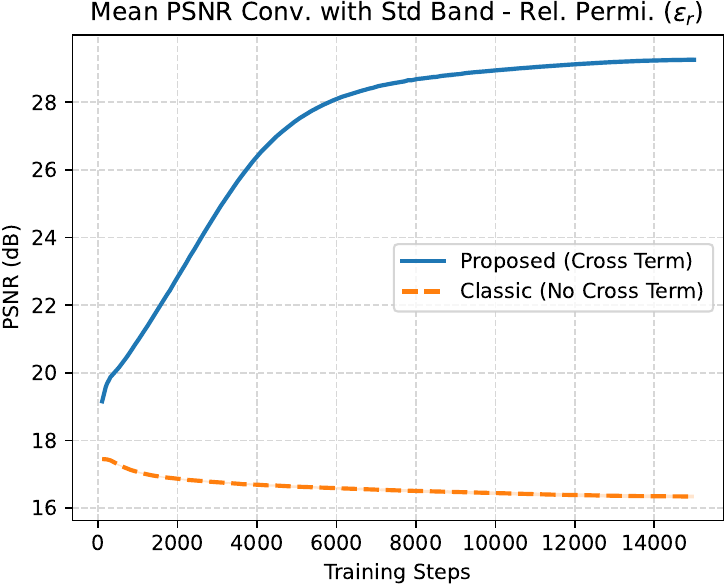}}\hspace*{\fill}
        
        \hspace*{\fill}%
        \subfloat[]{\includegraphics[width=0.4\linewidth]{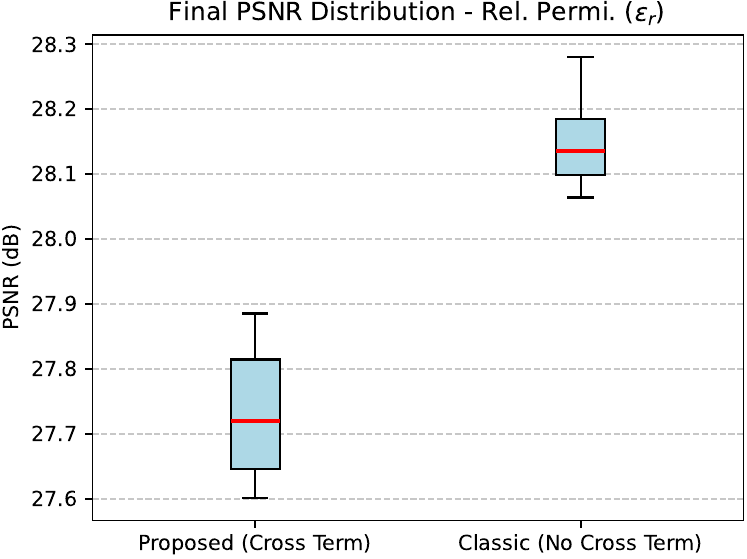}}\hfill
        \subfloat[]{\includegraphics[width=0.4\linewidth]{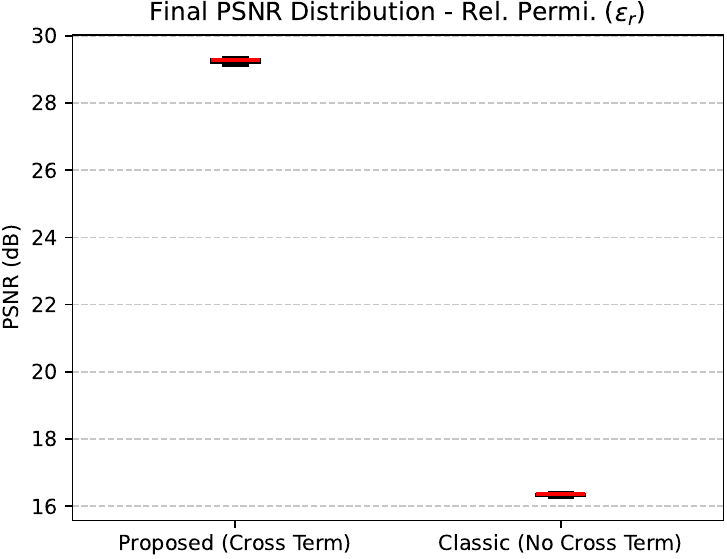}}\hspace*{\fill}
        
        \caption{Comparison of mean PSNR convergence with standard deviation band and boxplots for CC-PINN and classical cost function PINN using the simultaneous multi-frequency processing strategy to invert the Fresnel low-frequency dataset \textit{FoamTwinDielTM\_345} (a, c) and high-frequency dataset \textit{FoamTwinDielTM\_678} (b, d).}
        \label{fig:FoamTwinDielTM}   
    \end{figure}

    \begin{figure}[!t]
        \hspace*{\fill}%
        \subfloat[]{\includegraphics[height=0.30\linewidth]{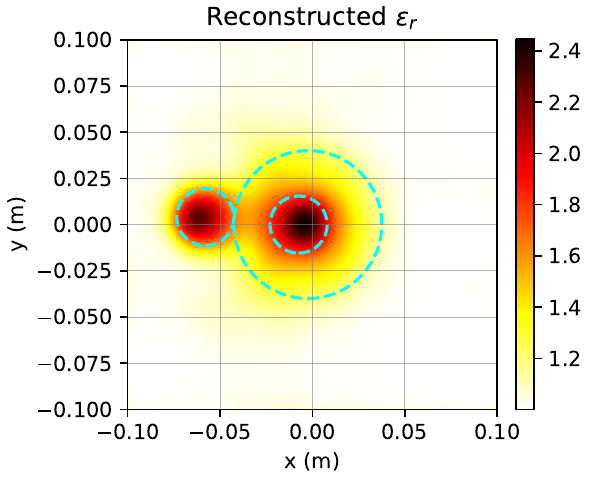}}\hfill
        \subfloat[]{\includegraphics[height=0.30\linewidth]{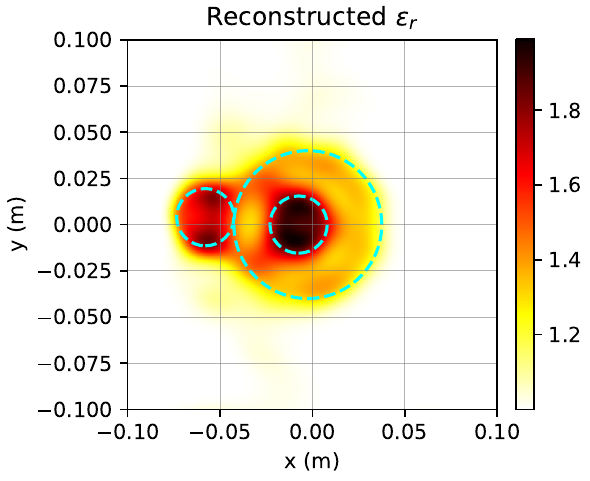}}\hspace*{\fill}

        \hspace*{\fill}%
        \subfloat[]{\includegraphics[height=0.30\linewidth]{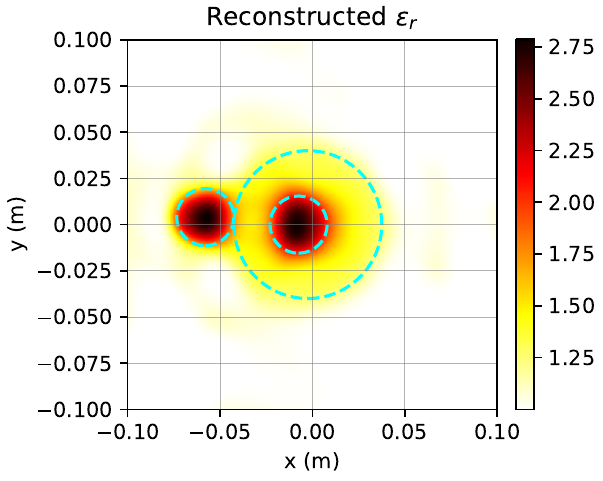}}\hfill
        \subfloat[]{\includegraphics[height=0.30\linewidth]{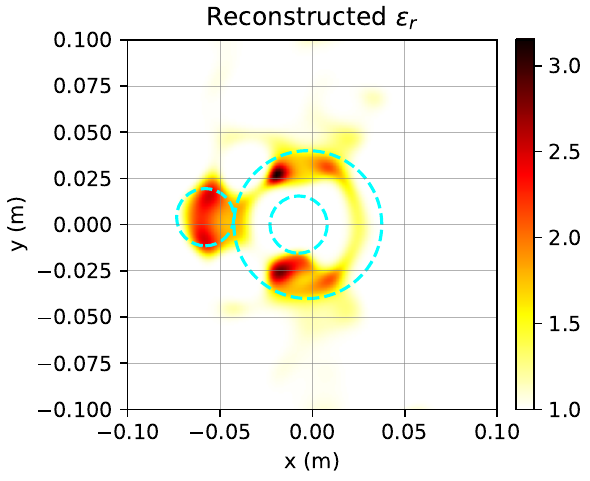}}\hspace*{\fill}
        \caption{Final reconstructed images for CC-PINN and classical cost function PINN using the frequency-hopping strategy to invert the Fresnel low-frequency dataset \textit{FoamTwinDielTM\_345} (a, c) and high-frequency dataset \textit{FoamTwinDielTM\_678} (b, d). Top: CC-PINN; Bottom: classical cost function PINN. The results shown are selected from the median of the final PSNR of the 11 independent runs. Dashed lines mark the true contours of the target. }
        \label{fig:FoamTwinDielTM_FH_recon}   
    \end{figure}

    \begin{figure}[!t]
        \hspace*{\fill}%
        \subfloat[]{\includegraphics[height=0.30\linewidth]{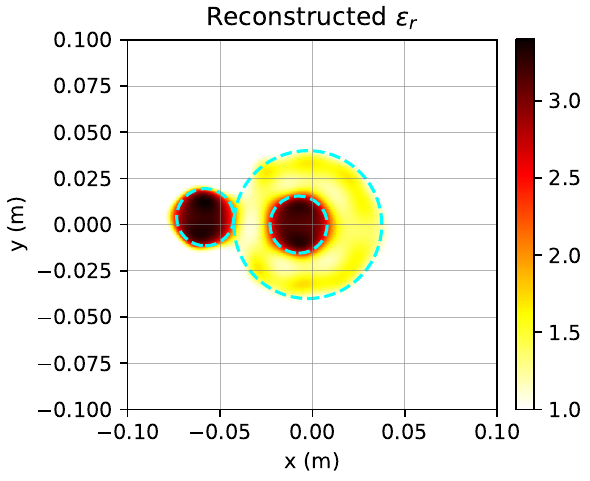}}\hfill
        \subfloat[]{\includegraphics[height=0.30\linewidth]{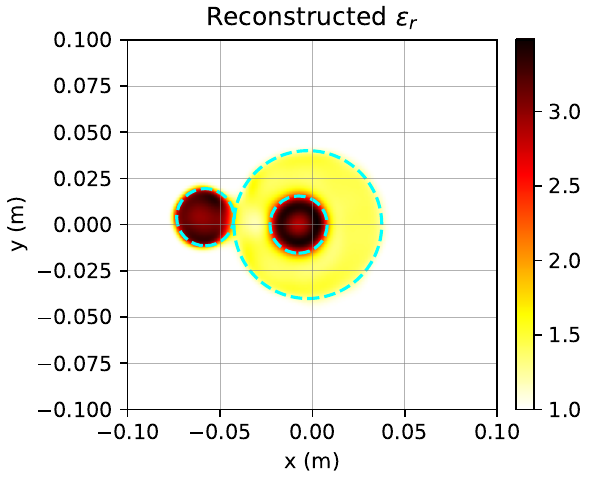}}\hspace*{\fill}

        \hspace*{\fill}%
        \subfloat[]{\includegraphics[height=0.30\linewidth]{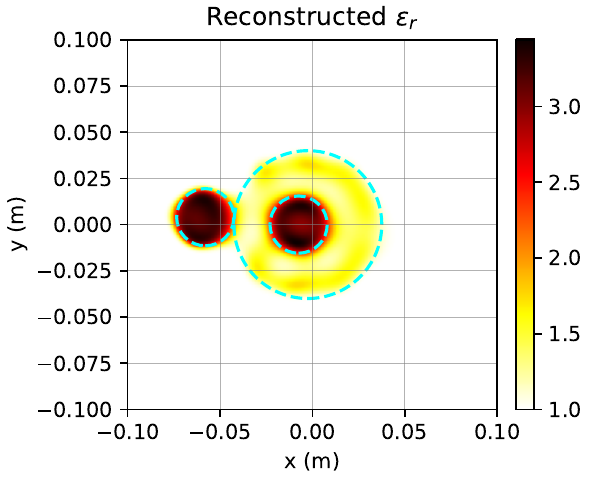}}\hfill
        \subfloat[]{\includegraphics[height=0.30\linewidth]{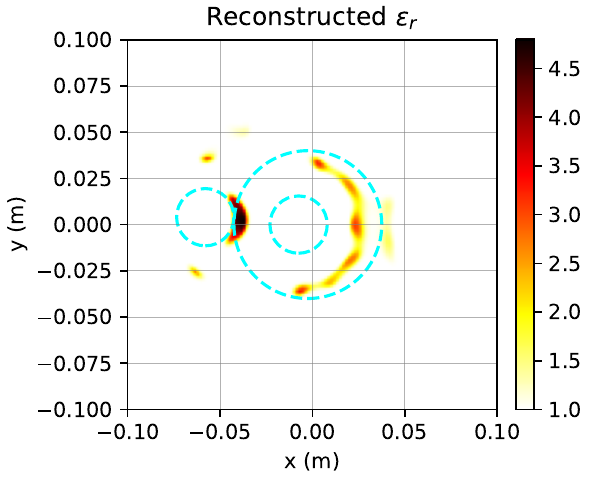}}\hspace*{\fill}
        \caption{Final reconstructed images for CC-PINN and classical cost function PINN using the simultaneous multi-frequency processing strategy to invert the Fresnel low-frequency dataset \textit{FoamTwinDielTM\_345} (a, c) and high-frequency dataset \textit{FoamTwinDielTM\_678} (b, d). Top: CC-PINN; Bottom: classical cost function PINN. The results shown are selected from the median of the final PSNR of the 11 independent runs. Dashed lines mark the true contours of the target. }
        \label{fig:FoamTwinDielTM_Sim_recon}   
    \end{figure}

    Both the frequency-hopping and simultaneous multi-frequency processing strategies are adopted to invert the low-frequency dataset ``FoamTwinDielTM\_345'' and the high-frequency dataset ``FoamTwinDielTM\_678'', with each group of experiments independently run 11 times and statistically analyzed. Fig.~\ref{fig:FoamTwinDielTM_FreqHp} and Fig.~\ref{fig:FoamTwinDielTM} respectively present the mean PSNR curves and boxplots for the two multi-frequency processing strategies on the two datasets. From the results, it can be seen that in measured data inversion, the simultaneous multi-frequency processing strategy is significantly superior to the frequency-hopping strategy. This is exactly the opposite of the simulation experiment results. As mentioned earlier, in the simulated data, the scatterers, incident field, and boundary conditions fully conform to the Helmholtz equation, and it is precisely this perfect physical consistency that causes sharp gradient conflicts. The measured Fresnel data contain factors such as non-ideal antenna patterns, coupling, wall reflections, and calibration residuals, none of which are modeled within the ``ideal equations'' of the PINN. If the frequency-hopping strategy is adopted, the PINN, in an attempt to forcibly satisfy the ``incorrect'' physical equations and fit these non-ideal data, will learn a pseudo-structure driven by systematic bias. When higher frequencies are added, the high frequencies are sensitive to geometric details and theoretically could correct the low-frequency errors, but at this point, the network parameters are already deeply trapped in the attraction basin of that pseudo-solution, making it very difficult for the high-frequency gradients to pull them out. In contrast, during simultaneous multi-frequency training, the loss function requires the same contrast distribution to simultaneously explain the scattered fields at multiple frequencies. The non-ideal factors in the measured data (such as random interference at a certain frequency, pattern jitter) are uncorrelated across different frequencies. The network is forced to find a solution that compromises and matches all frequencies, which instead creates an ensemble averaging effect, suppressing the tendency to overfit the measurement noise at a single frequency.

    Observing Fig.~\ref{fig:FoamTwinDielTM}, the PINN inversion algorithm using the traditional cost function converges significantly faster than CC-PINN when processing low-frequency data, but it fails to correctly invert the more difficult high-frequency data. CC-PINN can accurately converge to the ground truth for both the low-frequency dataset ``FoamTwinDielTM\_345'' and the high-frequency dataset ``FoamTwinDielTM\_678'', further verifying that CC-PINN possesses superior robustness compared to the PINN inversion algorithm employing the traditional cost function. Moreover, from the two boxplots in Fig.~\ref{fig:FoamTwinDielTM}, it can be seen that the final PSNR of CC-PINN on the high-frequency dataset is significantly higher than that on the low-frequency dataset, indicating that the weight-normalization-based Fourier feature network architecture proposed in this paper can accurately capture target characteristics from high-frequency information and is not significantly disturbed by measurement errors in high-frequency data. Fig.~\ref{fig:FoamTwinDielTM_FH_recon} and Fig~\ref{fig:FoamTwinDielTM_Sim_recon} present the inversion results of this set of measured data tests, with the results all corresponding to the median of the final PSNR from their respective 11 independent runs. The inversion results are obtained by inference using the saved PINN model. To better display the target details, the inference grid is refined to $128\times 128$, which precisely embodies a major advantage of the PINN inversion architecture: the PINN learns an implicit function mapping from spatial position to target contrast, which is essentially an infinite-dimensional continuous representation that can be queried at any spatial location, naturally adapting to irregular boundaries or multi-scale structures, and avoiding errors caused by grid discretization.

    Finally, this section presents the running times of CC-PINN for simulation and measured data experiments. The program code in this paper is built upon the PyTorch framework and runs on an NVIDIA A6000 GPU with CUDA acceleration. In the simulation experiments, taking a total training epoch number of 75000 as an example, a single experiment takes about 657 s under the frequency-hopping strategy and about 797 s under the simultaneous multi-frequency processing strategy; in the measured data experiments, with a total training epoch number of 15000, a single experiment takes about 167 s under the frequency-hopping strategy and about 179 s under the simultaneous multi-frequency processing strategy.

\section{Conclusion}\label{sec.Conclusion}

    This paper proposes a novel CC-PINN framework for solving inverse scattering problems. Based on a weight-normalized Fourier feature network architecture, and by introducing a cross-correlated cost function and a 2D-FFT fast computation model, the framework overcomes the bottlenecks of traditional PINNs, namely their tendency to fall into local optima and their computational inefficiency when solving electromagnetic inverse scattering problems. Experiments successfully demonstrate its robust reconstruction capability for high-contrast complex targets. It must be objectively pointed out here that although the reconstruction quality of the current CC-PINN has made a leap forward, its absolute robustness still slightly lags behind the traditional CC-CSI method based on rigorous analytical mathematical derivation when facing extremely ill-posed data or ultra-high contrasts. However, the PINN architecture based on deep learning possesses unique advantages that traditional deterministic algorithms cannot match, representing an important direction for future development: 1) Continuous super-resolution representation: PINN uses a neural network to accomplish the continuous mapping from spatial coordinates to dielectric parameters, naturally breaking free from the constraints of traditional grid discretization and possessing the potential to achieve ``mesh-free'' super-resolution imaging; 2) Extremely flexible prior fusion: introducing target prior constraints into a PINN merely requires adding simple regularization terms to the loss function, without the need to derive complex operators as in traditional algorithms; 3) Multi-physics coupling expansion capability: the PINN architecture can be easily extended to thermo-electromagnetic coupled inversion or acousto-electromagnetic joint imaging by sharing hidden layers or merging PDE residuals. 

    In future research, we will further explore implicitly embedding the prior knowledge of traditional physical operators into the neural network structure, attempting to bridge and ultimately surpass traditional nonlinear inverse scattering algorithms in terms of algorithmic robustness. The sustained research on traditional nonlinear inverse scattering algorithms will inevitably continue to inspire the in-depth development of neural networks in terms of physical mechanisms and other aspects.

\ifCLASSOPTIONcaptionsoff
  \newpage
\fi

\bibliographystyle{ieeetr}
\bibliography{mybib}

@incollection{hajebi2024subsurface,
  title={Subsurface Inverse Profiling and Imaging Using Stochastic Optimization Techniques},
  author={Hajebi, Maryam and Hoorfar, Ahmad},
  booktitle={Signal and Image Processing for Remote Sensing},
  pages={45--75},
  year={2024},
  publisher={CRC Press}
}

@ARTICLE{HajebiTGRS2022,
  author={Hajebi, Maryam and Hoorfar, Ahmad},
  journal={IEEE Transactions on Geoscience and Remote Sensing}, 
  title={Multiple Buried Target Reconstruction Using a Multiscale Hybrid of Diffraction Tomography and {CMA-ES} Optimization}, 
  year={2022},
  volume={60},
  number={},
  pages={1-13},
  doi={10.1109/TGRS.2022.3206722}}

@ARTICLE{PastorinoTAP2007,
  author={Pastorino, Matteo},
  journal={IEEE Transactions on Antennas and Propagation}, 
  title={Stochastic Optimization Methods Applied to Microwave Imaging: A Review}, 
  year={2007},
  volume={55},
  number={3},
  pages={538-548},
  doi={10.1109/TAP.2007.891568}}

@ARTICLE{DastfalTGRSL2025,
  author={Dastfal, Zahra and Hajebi, Maryam and Sharifzadeh, Mansoureh and Hoorfar, Ahmad},
  journal={IEEE Geoscience and Remote Sensing Letters}, 
  title={Learning-Based Profiling of Buried Elliptical-Cylindrical Objects}, 
  year={2025},
  volume={22},
  number={},
  pages={1-5},
  doi={10.1109/LGRS.2025.3543290}}

@ARTICLE{851784,
  author={Tie Jun Cui and Weng Cho Chew},
  journal={IEEE Transactions on Geoscience and Remote Sensing}, 
  title={Novel diffraction tomographic algorithm for imaging two-dimensional targets buried under a lossy Earth}, 
  year={2000},
  volume={38},
  number={4},
  pages={2033-2041},
  doi={10.1109/36.851784}}

@ARTICLE{10144396,
  author={Wang, Miao and Sun, Shilong and Dai, Dahai and Su, Yi and Wu, Manqing},
  journal={IEEE Transactions on Geoscience and Remote Sensing}, 
  title={Quantitative Diffraction Tomography for Weak Scatterers Based on Aliasing Modification of the Multifrequency Spatial Spectrum}, 
  year={2023},
  volume={61},
  number={},
  pages={1-14},
  doi={10.1109/TGRS.2023.3282937}}

@article{Tarantola1984,
  author  = {Tarantola, Albert},
  title   = {Inversion of Seismic Reflection Data in the Acoustic Approximation},
  journal = {Geophysics},
  year    = {1984},
  volume  = {49},
  number  = {8},
  pages   = {1259--1266},
  doi     = {10.1190/1.1441754}
}

@article{litman1998reconstruction,
  title={Reconstruction of a two-dimensional binary obstacle by controlled evolution of a level-set},
  author={Litman, Amelie and Lesselier, Dominique and Santosa, Fadil},
  journal={Inverse Problems},
  volume={14},
  number={3},
  pages={685-706},
  year={1998},
  publisher={IOP Publishing}
}

@article{van2001contrast,
  title={Contrast source inversion method: state of art},
  author={van den Berg, P. M. and Abubakar, A},
  journal={Journal of Electromagnetic Waves and Applications},
  volume={15},
  number={11},
  pages={1503--1505},
  year={2001}
}

@inproceedings{belkebir1996using,
  title={Using multiple frequency information in the iterative solution of a two-dimensional nonlinear inverse problem},
  author={Belkebir, K and Tijhuis, AG},
  booktitle={Proceedings Progress in Electromagnetics Research Symposium, PIERS 1996, 8 July 1996, Innsbruck, Germany},
  pages={353},
  year={1996},
  organization={University of Innsbruck}
}

@article{Isaacson1986,
  author  = {Isaacson, David},
  title   = {Distinguishability of Conductivities by Electric Current Computed Tomography},
  journal = {IEEE Transactions on Medical Imaging},
  year    = {1986},
  volume  = {5},
  number  = {2},
  pages   = {91--95},
  doi     = {10.1109/TMI.1986.4307752}
}

@article{Seo2003,
  author  = {Seo, Jin Keun and Woo, Eung Je},
  title   = {Nonlinear Inverse Problems in Imaging},
  journal = {Annual Review of Biomedical Engineering},
  year    = {2003},
  volume  = {5},
  pages   = {413--449},
  doi     = {10.1146/annurev.bioeng.5.040202.121725}
}

@article{Cheney2000,
  author  = {Cheney, Margaret and Isaacson, David and Newell, Jonathan C.},
  title   = {Electrical Impedance Tomography},
  journal = {SIAM Review},
  year    = {2000},
  volume  = {41},
  number  = {1},
  pages   = {85--101},
  doi     = {10.1137/S0036144598333613}
}

@article{Schuster2017,
  author  = {Schuster, Gerard T.},
  title   = {Seismic Inversion},
  journal = {Society of Exploration Geophysicists},
  year    = {2017},
  doi     = {10.1190/1.9781560803423}
}

@article{sun2017cross,
  title={Cross-correlated contrast source inversion},
  author={Sun, Shilong and Kooij, Bert Jan and Jin, Tian and Yarovoy, Alexander G},
  journal={IEEE Transactions on Antennas and Propagation},
  volume={65},
  number={5},
  pages={2592--2603},
  year={2017},
  publisher={IEEE}
}

@article{sun2018inversion,
  title={Inversion of multifrequency data with the cross-correlated contrast source inversion method},
  author={Sun, Shilong and Kooij, Bert-Jan and Yarovoy, Alexander G},
  journal={Radio Science},
  volume={53},
  number={6},
  pages={710--723},
  year={2018},
  publisher={AGU}
}

@article{geffrin2005free,
  title={Free space experimental scattering database continuation: experimental set-up and measurement precision},
  author={Geffrin, Jean-Michel and Sabouroux, Pierre and Eyraud, Christelle},
  journal={Inverse Problems},
  volume={21},
  number={6},
  pages={S117--S130},
  year={2005},
  publisher={IOP Publishing}
}

@TECHREPORT{Malcolm_Slaney,
  AUTHOR =        {Malcolm Slaney and A. C. Kak},
  TITLE =         {Imaging with Diffraction Tomography},
  NUMBER =        {TR-EE 85-5},
  INSTITUTION =   {Department of Electrical and Computer Engineering, Purdue University},
  ADDRESS =       {West Lafayette, Indiana 47907},
  NOTE =          {},
  MONTH =         {February},
  YEAR  =         {1985},
  PAGES =         {3},
  URL   =         {https://docs.lib.purdue.edu/ecetr/540},
  CONTACT =       {epubs@purdue.edu}
}

@ARTICLE{5210141,
  author={Chen, Xudong},
  journal={IEEE Transactions on Geoscience and Remote Sensing}, 
  title={Subspace-Based Optimization Method for Solving Inverse-Scattering Problems}, 
  year={2010},
  volume={48},
  number={1},
  pages={42-49},
  doi={10.1109/TGRS.2009.2025122}
  }

@ARTICLE{10663369,
  author={Wang, Miao and Sun, Shilong and Dai, Dahai and Zhang, Yongsheng and Su, Yi},
  journal={IEEE Transactions on Antennas and Propagation}, 
  title={Cross-Correlated Subspace-Based Optimization Method for Solving Electromagnetic Inverse Scattering Problems}, 
  year={2024},
  volume={72},
  number={11},
  pages={8575-8589},
  doi={10.1109/TAP.2024.3450328}
  }

@article{Li2010,
  author  = {Li, Jing and Abubakar, Aria and van den Berg, Peter M.},
  title   = {A Subspace-Based Optimization Method for Solving Inverse Scattering Problems},
  journal = {IEEE Transactions on Geoscience and Remote Sensing},
  year    = {2010},
  volume  = {48},
  number  = {1},
  pages   = {42--49},
  doi     = {10.1109/TGRS.2009.2027690}
}

@book{colton2012inverse,
  title={Inverse acoustic and electromagnetic scattering theory},
  author={Colton, David and Kress, Rainer},
  volume={93},
  year={2012},
  publisher={Springer Science \& Business Media}
}

@article{van1997contrast,
  title={A contrast source inversion method},
  author={Van Den Berg, Peter M and Kleinman, Ralph E},
  journal={Inverse Problems},
  volume={13},
  number={6},
  pages={1607--1620},
  year={1997},
  publisher={IOP Publishing}
}

@PHDTHESIS{W.Shin2013,
  author = {Wonseok Shin},
  title = {3{D} finite-difference frequency-domain method for plasmonics and nanophotonics},
  school = {Stanford University},
  year = {2013}
}

@inproceedings{NEURIPS2020_55053683,
 author = {Tancik, Matthew and Srinivasan, Pratul and Mildenhall, Ben and Fridovich-Keil, Sara and Raghavan, Nithin and Singhal, Utkarsh and Ramamoorthi, Ravi and Barron, Jonathan and Ng, Ren},
 booktitle = {Advances in Neural Information Processing Systems},
 editor = {H. Larochelle and M. Ranzato and R. Hadsell and M.F. Balcan and H. Lin},
 pages = {7537--7547},
 publisher = {Curran Associates, Inc.},
 title = {Fourier Features Let Networks Learn High Frequency Functions in Low Dimensional Domains},
 url = {https://proceedings.neurips.cc/paper_files/paper/2020/file/55053683268957697aa39fba6f231c68-Paper.pdf},
 volume = {33},
 year = {2020}
}

@inproceedings{DBLP:conf/nips/RahimiR07,
  author       = {Ali Rahimi and
                  Benjamin Recht},
  editor       = {John C. Platt and
                  Daphne Koller and
                  Yoram Singer and
                  Sam T. Roweis},
  title        = {Random Features for Large-Scale Kernel Machines},
  booktitle    = {Advances in Neural Information Processing Systems 20, Proceedings
                  of the Twenty-First Annual Conference on Neural Information Processing
                  Systems, Vancouver, British Columbia, Canada, December 3-6, 2007},
  pages        = {1177--1184},
  publisher    = {Curran Associates, Inc.},
  year         = {2007},
  url          = {https://proceedings.neurips.cc/paper/2007/hash/013a006f03dbc5392effeb8f18fda755-Abstract.html},
  bibsource    = {dblp computer science bibliography, https://dblp.org}
}

@inproceedings{10.5555/3157096.3157197,
author = {Salimans, Tim and Kingma, Diederik P.},
title = {Weight normalization: a simple reparameterization to accelerate training of deep neural networks},
year = {2016},
isbn = {9781510838819},
publisher = {Curran Associates Inc.},
address = {Red Hook, NY, USA},
booktitle = {Proceedings of the 30th International Conference on Neural Information Processing Systems},
pages = {901–909},
numpages = {9},
location = {Barcelona, Spain},
series = {NIPS'16}
}

@ARTICLE{11106328,
  author={Sun, Shilong},
  journal={IEEE Transactions on Antennas and Propagation}, 
  title={Frequency-Binned Cumulative Hopping Framework Incorporating Wavelength-Dependent Weighting Strategies for Inverse Scattering of High-Contrast Objects}, 
  year={2025},
  volume={73},
  number={10},
  pages={8048-8062},
  doi={10.1109/TAP.2025.3592795}}

@ARTICLE{10044704,
  author={Sun, Shilong and Dai, Dahai and Wang, Xuesong},
  journal={IEEE Transactions on Antennas and Propagation}, 
  title={A Fast Algorithm of Cross-Correlated Contrast Source Inversion in Homogeneous Background Media}, 
  year={2023},
  volume={71},
  number={5},
  pages={4380-4393},
  doi={10.1109/TAP.2023.3243768}}

@article{van2003multiplicative,
  title={Multiplicative regularization for contrast profile inversion},
  author={van den Berg, Peter M and Abubakar, Aria and Fokkema, Jacob T},
  journal={Radio Science},
  volume={38},
  number={2},
  year={2003},
  publisher={Wiley Online Library}
}

@article{chew1990reconstruction,
  title={Reconstruction of two-dimensional permittivity distribution using the distorted {B}orn iterative method},
  author={Chew, Weng Cho and Wang, Yi-Ming},
  journal={IEEE Transactions on Medical Imaging},
  volume={9},
  number={2},
  pages={218--225},
  year={1990},
  publisher={IEEE}
}

@ARTICLE{10028725,
  author={Zhang, Huan Huan and Yao, He Ming and Jiang, Lijun and Ng, Michael},
  journal={IEEE Transactions on Antennas and Propagation}, 
  title={Solving Electromagnetic Inverse Scattering Problems in Inhomogeneous Media by Deep Convolutional Encoder–Decoder Structure}, 
  year={2023},
  volume={71},
  number={3},
  pages={2867-2872},
  doi={10.1109/TAP.2023.3239185}}

@ARTICLE{10666745,
  author={Wang, Miao and Sun, Shilong and Zhang, Yongsheng and Dai, Dahai and Wu, Hao and Su, Yi},
  journal={IEEE Transactions on Microwave Theory and Techniques}, 
  title={{PUP-Net}: A Twofold Physical Model Embedded {3-D} {U-Net} With Polarization Fusion for Solving Inverse Scattering Problems With a Sparse Planar Array}, 
  year={2025},
  volume={73},
  number={4},
  pages={2123-2136},
  doi={10.1109/TMTT.2024.3450684}}

@ARTICLE{752198,
  author={Caorsi, S. and Gamba, P.},
  journal={IEEE Transactions on Geoscience and Remote Sensing}, 
  title={Electromagnetic detection of dielectric cylinders by a neural network approach}, 
  year={1999},
  volume={37},
  number={2},
  pages={820-827},
  doi={10.1109/36.752198}}

@ARTICLE{8565987,
  author={Li, Lianlin and Wang, Long Gang and Teixeira, Fernando L. and Liu, Che and Nehorai, Arye and Cui, Tie Jun},
  journal={IEEE Transactions on Antennas and Propagation}, 
  title={DeepNIS: Deep Neural Network for Nonlinear Electromagnetic Inverse Scattering}, 
  year={2019},
  volume={67},
  number={3},
  pages={1819-1825},
  doi={10.1109/TAP.2018.2885437}}

@ARTICLE{9897092,
  author={Zhang, Hongrui and Chen, Yanjin and Cui, Tie Jun and Teixeira, Fernando L. and Li, Lianlin},
  journal={IEEE Transactions on Microwave Theory and Techniques}, 
  title={Probabilistic Deep Learning Solutions to Electromagnetic Inverse Scattering Problems Using Conditional Renormalization Group Flow}, 
  year={2022},
  volume={70},
  number={11},
  pages={4955-4965},
  doi={10.1109/TMTT.2022.3205890}}

@ARTICLE{9852109,
  author={Guo, Rui and Lin, Zhichao and Li, Maokun and Yang, Fan and Xu, Shenheng and Abubakar, Aria},
  journal={IEEE Transactions on Antennas and Propagation}, 
  title={A Nonlinear Model Compression Scheme Based on Variational Autoencoder for Microwave Data Inversion}, 
  year={2022},
  volume={70},
  number={11},
  pages={11059-11069},
  doi={10.1109/TAP.2022.3195553}}

@ARTICLE{10572305,
  author={Du, Naike and Wang, Jing and Song, Rencheng and Xu, Kuiwen and Sun, Sheng and Ye, Xiuzhu},
  journal={IEEE Transactions on Microwave Theory and Techniques}, 
  title={Inhomogeneous Media Inverse Scattering Problem Assisted by Swin Transformer Network}, 
  year={2024},
  volume={72},
  number={12},
  pages={6809-6820},
  doi={10.1109/TMTT.2024.3412113}}

@article{YuSun18,
  author = {Yu Sun and Zhihao Xia and Ulugbek S. Kamilov},
  journal = {Opt. Express},
  number = {11},
  pages = {14678--14688},
  publisher = {Optica Publishing Group},
  title = {Efficient and accurate inversion of multiple scattering with deep learning},
  volume = {26},
  month = {May},
  year = {2018},
  url = {https://opg.optica.org/oe/abstract.cfm?URI=oe-26-11-14678},
  doi = {10.1364/OE.26.014678},
}

@article{RAISSI2019686,
title = {Physics-informed neural networks: A deep learning framework for solving forward and inverse problems involving nonlinear partial differential equations},
journal = {Journal of Computational Physics},
volume = {378},
pages = {686-707},
year = {2019},
issn = {0021-9991},
doi = {https://doi.org/10.1016/j.jcp.2018.10.045},
url = {https://www.sciencedirect.com/science/article/pii/S0021999118307125},
author = {M. Raissi and P. Perdikaris and G.E. Karniadakis}
}

@article{Chen2019PhysicsinformedNN,
  title={Physics-informed neural networks for inverse problems in nano-optics and metamaterials.},
  author={Yuyao Chen and Lu Lu and George Em Karniadakis and Luca Dal Negro},
  journal={Optics express},
  year={2019},
  volume={28 8},
  pages={
          11618-11633
        },
  url={https://api.semanticscholar.org/CorpusID:208547648}
}




\end{document}